\documentclass[journal]{IEEEtran} 

\usepackage{graphicx,amsmath,amssymb,cite,citesort,subfigure,bm,color,algorithm,algorithmic,} 

\usepackage{multicol,multirow}
\usepackage{url}
\usepackage[T1]{fontenc}

\begin{document}

\title{
  Single-Carrier Delay-Doppler Domain Equalization
}

\author{
  \authorblockN{
    Yuto Hama\authorrefmark{1}, \IEEEmembership{Member, IEEE,} and Hideki Ochiai\authorrefmark{2}, \IEEEmembership{Fellow, IEEE}
  }\\
  \authorblockA{
    Department of Electrical and Computer Engineering,\\
    Yokohama National University\\
    79-5 Tokiwadai, Hodogaya, Yokohama, Kanagawa 240-8501, Japan\\
    Email: {\authorrefmark{1}}yuto.hama@ieee.org},
  \authorblockA{
    \authorrefmark{2}hideki.ochiai@ieee.org 
  }
  \vspace{-0.5cm}
}

\maketitle

\begin{abstract}
  For doubly-selective channels,
  delay-Doppler~(DD) modulation,
  mostly known as orthogonal time frequency space~(OTFS) modulation,
  enables simultaneous compensation of delay and Doppler shifts.
  However,
  OTFS modulated signal has high peak-to-average power ratio~(PAPR)
  because of its precoding operation performed over the DD domain.
  In order to deal with this problem,
  we propose a single-carrier transmission
  with delay-Doppler domain equalization~(SC-DDE).
  In this system,
  the discretized time-domain SC signal is converted to the DD domain
  by discrete Zak transform~(DZT) at the receiver side,
  followed by delay-Doppler domain equalization~(DDE).
  Since equalization is performed in the DD domain,
  the SC-DDE receiver should acquire the channel delay-Doppler response.
  To this end,
  we introduce an embedded pilot-aided channel estimation scheme designed for SC-DDE,
  which does not affect the peak power property of transmitted signals.
  Through computer simulation,
  distribution of PAPR and bit error rate~(BER) performance of the proposed system
  are compared with those of the conventional OTFS and SC with frequency-domain equalization~(SC-FDE).
  As a result,
  our proposed SC-DDE significantly outperforms SC-FDE in terms of BER
  at the expense of additional computational complexity at the receiver.
  Furthermore,
  SC-DDE shows much lower PAPR than OTFS
  even though they achieve comparable coded BER performance.
\end{abstract}



\IEEEpeerreviewmaketitle

\section{Introduction}
\label{sec:introduction}

In mobile wireless communications,
the system performance is affected by time- and/or frequency-selective fading
caused by multipath channels.
In most cases,
frequency-selective fading is not avoidable
since the propagation delay occurs frequently
in practical environments due to reflection and diffraction.
In orthogonal frequency-division multiplexing~(OFDM) signaling,
the interference caused by frequency-selective fading can be avoided 
provided that
a cyclic prefix~(CP) is longer than the multipath delay.
Therefore,
OFDM is a {\it de-fact} standard
in modern wireless communication standards
such as cellular networks~(4G and 5G) and Wi-Fi~\cite{dahlman20134g}.

As a main drawback of OFDM,
it exhibits a high peak-to-average power ratio (PAPR),
which leads to low power amplifier~(PA) efficiency~\cite{ochiai2001distribution}.
Furthermore,
OFDM exhibits susceptibility to time-selective fading induced by Doppler shifts~\cite{wang2006performance}.
The former issue is addressed by discrete Fourier transform~(DFT)-precoding,
which in fact transforms the transmit signal into
single-carrier~(SC) equivalence~\cite{ochiai_instantaneous_2012}. 
Thus,
it is OFDM-based SC implementation known as DFT-spread OFDM~(DFT-s-OFDM),
which is adopted in the uplink of 4G and 5G~\cite{myung2006single},
and also envisioned in 6G~\cite{rikkinen2020thz}.
At the receiver,
SC with frequency domain equalization~(SC-FDE) is employed
so as to mitigate the effect of frequency-selective fading~\cite{falconer2002frequency}.
However,
it is not designed for time-selective fading caused by Doppler shifts.

In order to cope with doubly- (time and frequency) selective fading channels,
delay-Doppler modulation schemes,
such as orthogonal time frequency space~(OTFS) modulation~\cite{hadani2017orthogonal}
and orthogonal delay-Doppler division multiplexing~(ODDM) modulation~\cite{lin2022orthogonal},
have been proposed and actively investigated in recent years~\cite{wei2021orthogonal,li2021performance,lin2023multi}.
OTFS is a two-dimensional~(2D) modulation technique,
and modulated symbols are mapped onto the delay-Doppler~(DD) domain.
It thus enables efficient equalization for doubly-selective fading performed in the DD domain.
Several implementation approaches have been investigated for OTFS modulation~\cite{raviteja2018practical,yuan2021iterative,lampel2022otfs},
and these are now considered as a part of delay-Doppler modulation~\cite{lin2023multi}.
Among them,
we focus on the widely investigated OFDM-based OTFS with reduced CP
due to its practicality and compatibility~\cite{raviteja2018practical},
which we consistently refer 
to as OTFS throughout this work.
Even though 
OTFS is an attractive scheme to  
cope with Doppler shift,
its high signal PAPR is a major practical issue similar to OFDM systems~\cite{surabhi2019peak,wei2021charactering}.
Thus,
several PAPR reduction approaches have been proposed for OTFS systems~\cite{gao2020peak,naveen2020peak}.
Nevertheless, 
these approaches require additional computational complexity at the transmitter
and also cause the degradation of the error rate performance.

Based on the above observation,
in this paper,
we propose delay-Doppler domain equalization~(DDE) for SC signaling, 
which we call SC-DDE,
where the entire compensation process of OTFS modulation
is placed at the receiver side.
The SC-DDE receiver converts the received time domain symbols into the DD domain
by the discrete Zak transform~(DZT)~\cite{lampel2022otfs}.
Then, DDE is performed based on minimum mean square error~(MMSE).
Since the transmitter in our SC-DDE system is equivalent to the conventional SC transmission with CP,
it is suitable for uplink communications
similar to the conventional SC-FDE.
As a major challenge of SC-DDE as well as OTFS,
the receiver must have the channel state information~(CSI) in the DD domain,
which corresponds to the channel delay-Doppler response.
In order to deal with this problem,
we introduce an embedded pilot-aided channel estimation scheme designed for SC-DDE,
which does not affect the peak power property of SC transmission.
Through computer simulation,
we show that the resulting PAPR of SC-DDE is significantly lower than that of OTFS
even in the presence of pilot symbols for channel estimation.
Furthermore,
we demonstrate that
the proposed SC-DDE significantly outperforms SC-FDE
and achieves comparable performance to OTFS
in terms of coded bit error rate~(BER) over doubly-selective fading channels.

The primary contributions of this paper are summarized as follows:
\begin{itemize}
\item We propose an SC-DDE system that can compensate for doubly-selective fading caused by delay and Doppler shift while retaining the low PAPR property of SC transmission.
\item We introduce an embedded pilot-aided channel estimation scheme for SC-DDE that retains the low peak power property of SC signals.
\item We demonstrate that our proposed SC-DDE significantly outperforms the conventional SC-FDE over doubly-selective channels. Furthermore, the coded BER performance of SC-DDE is comparable to that of OTFS while enjoying much lower PAPR.
\end{itemize}

This paper is organized as follows.
The SC transmission over doubly-selective channels is described in Section~\ref{sec:sc_transmission}.
Section~\ref{sec:sc_dde} introduces the proposed single-carrier delay-Doppler equalization~(SC-DDE),
and its computational complexity is compared with existing systems.
The embedded pilot-aided channel estimation for SC-DDE is proposed in Section~\ref{sec:channel_estimation}.
The simulation results are shown in Section~\ref{sec:simulation},
where PAPR distribution and BER performance are compared.
Section~\ref{sec:conclusion} concludes this work.

\subsubsection*{Notation}
Throughout this paper,
the $k$th element of a vector~$\mathbf{a}$ is denoted by
either ${a}_k$ or $\mathbf{a}[k]$.
Similarly,
both ${A}_{l,k}$ and $\mathbf{A}[l,k]$ denote the $(l,k)$th element of a matrix~$\mathbf{A}$.

\section{Single-Carrier Transmission Over Doubly-Selective Channels}
\label{sec:sc_transmission}

\begin{figure}[tb]
  \centering
  \includegraphics[width = \hsize, clip]{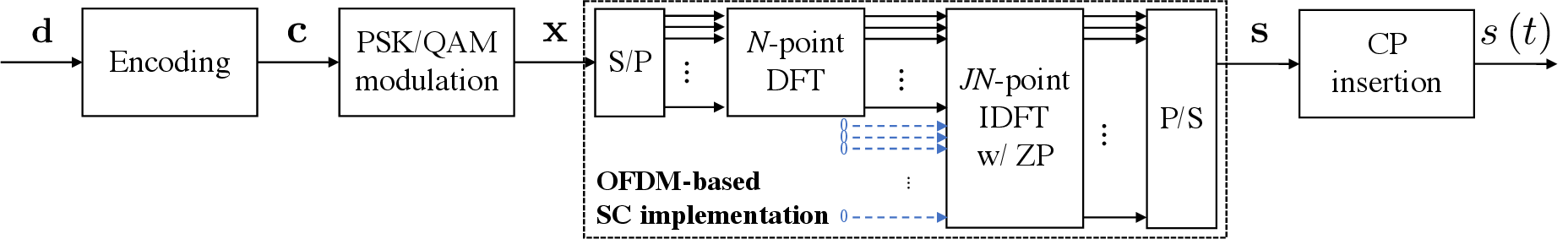}
  \caption{A block diagram of the SC transmitter 
    with $J$-times oversampling for the block length~$N$.}
  \label{transmitter}
\end{figure}

We describe a single-carrier transmission over doubly-selective channels.
In this work,
we focus on the block transmission with SC signal
generated by OFDM with DFT-precoding 
rather than the conventional pulse-shaping-based system,
since the former is adopted in the uplink of modern cellular network standards (4G and 5G).
The block diagram of the SC transmitter is shown in Fig.~\ref{transmitter}.

\subsection{Single-Carrier Transmission}
\label{sec:sc}

The transmitter employs the SC block transmission with CP
as in the conventional SC-FDE systems.
First,
the information bit sequence~$\mathbf{d} \in \left\{ 0,1 \right\}^{D}$
is encoded by a binary error correcting code with code rate~$R_c$,
and the encoded bit sequence is given by~$\mathbf{c} \in \left\{ 0,1 \right\}^{D/R_c}$.
Then,
PSK or QAM modulation is performed
for the encoded bit sequence.
Let $x_n \in \mathcal{X}$ denote the $n$th modulated symbol
with its average power given by~$E_s$,
where $\mathcal{X}$ is a set of the signal constellation points.
The modulated symbol vector of length~$N$ is given by~$\mathbf{x} = \left( x_0, \cdots, x_{N-1} \right)^T \in \mathcal{X}^{N\times1}$,
where $N$ corresponds to the block length.
The modulated symbol vector~$\mathbf{x}$ is spread over the frequency domain by DFT-precoding as\footnote{Throughout this paper, the superscripts of $\mathcal{F}$ and $\mathcal{D}$ indicate that the symbols $a^\mathcal{F}$ and $a^\mathcal{D}$ are represented in the frequency and delay-Doppler domain, respectively.}
\begin{align}
  \mathbf{x}^\mathcal{F} 
  &=
  \mathbf{F}_{N} \mathbf{x},
  \label{eq:x_FD}
\end{align}
where $\mathbf{F}_{N}$ denotes the DFT matrix of size~$N$.
By performing the IDFT with zero padding~(ZP),
the resulting SC transmitted symbol vector with $J$-times oversampling,
denoted by~$\mathbf{s} = \left( s_0, \cdots, s_{JN-1}  \right)^T \in \mathbb{C}^{JN \times 1}$,
is expressed as~\cite{hama2022noma}
\begin{align}
  \mathbf{s}
  &=
  \mathbf{F}_{N,J}^\text{H}
  \mathbf{x}^\mathcal{F} 
  =
  \mathbf{F}_{N,J}^\text{H}
  \mathbf{F}_{N} \mathbf{x}
  ,
  \label{eq:s}
\end{align}
with
\begin{align}
  \mathbf{F}_{N,J}
  \triangleq
  \begin{bmatrix}
    \mathbf{I}_{N} & \mathbf{O}_{N \times (J-1)N}
  \end{bmatrix}
  \mathbf{F}_{JN}
  ,
\end{align}
where $\mathbf{X}^\text{H}$ represents the Hermitian transpose of a matrix~$\mathbf{X}$,
$\mathbf{I}_N$ is the identity matrix of size~$N$,
and $\mathbf{O}_{M\times N}$ is the zero matrix of size~$M\times N$.

Let $s(t)$ denote the continuous-time transmitted SC signal
with the Nyquist interval~$T_0$,
where the effective block time duration is given by~$T_\text{block} \triangleq N T_0$.
From~\eqref{eq:s},
the $m$th sample of the transmitted symbol vector with $J$-times oversampling
is given by~\cite{ochiai_instantaneous_2012}
\begin{align}
  s_m
  &
  \triangleq
  s \left( m \frac{T_0}{J} \right)
  \nonumber\\&
  =
  \sum_{n=0}^{N-1}
  x_n
  e^{j\pi \left( 1 - \frac{1}{N} \right) \left( \frac{m}{J}-n \right)}
  \frac{\sin \left\{ \pi \left( \frac{m}{J}-n \right) \right\}}{N \sin \left\{ \frac{\pi}{N} \left( \frac{m}{J}-n \right) \right\}}
  ,
  \nonumber\\&
  \qquad\qquad\qquad
  m \in \left\{ 0, \cdots, JN-1 \right\}.
  \label{eq:s_m}
\end{align}
Prior to transmission,
CP with length~$N_\text{CP}$ is inserted
so as to compensate for the inter-block interference~(IBI).
In this work, as is the case with practical OFDM-based cellular network systems~\cite{hama2022noma},
we assume that the CP length~$N_\text{CP}$ is sufficient such that the effect of IBI is negligible.
As a result,
the continuous-time SC signal 
is expressed as~\cite{ochiai_instantaneous_2012}
\begin{align}
  s \left( t \right)
  &=
  \sum_{n=-N_\text{CP}}^{N-1}
  x_{[n]_N}
  g_\text{psinc} \left( t - n T_0 \right)
  w_{[-N_\text{CP} T_0, NT_0]}
  ,
  \label{eq:sc_ofdm}
\end{align}
where $[ n ]_N \triangleq n \bmod N \in \{ 0, \cdots, N-1 \}$ represents modulo operation.
In~\eqref{eq:sc_ofdm},
$g_\text{psinc} \left( t \right)$ denotes a periodic version (with period~$N$) of the sinc pulse
defined as~\cite{ochiai_instantaneous_2012}
\begin{align}
  g_\text{psinc} \left( t \right)
  &\triangleq
  e^{j\pi \left( 1 - \frac{1}{N} \right) \frac{t}{T_0}}
  \frac{\sin \left( \pi \frac{t}{T_0} \right)}{N \sin \left( \frac{\pi}{N} \frac{t}{T_0} \right)}
  ,
  \label{eq:psinc}
\end{align}
and
$w_{[T_\text{low}, T_\text{high}]} \left( t \right)$ is a rectangular window function defined as
\begin{align}
  w_{[T_\text{low}, T_\text{high}]} \left( t \right)
  &\triangleq
  \begin{cases}
    1, & T_\text{low} \leq t \leq T_\text{high}, \\
    0, & \text{otherwise}.
  \end{cases}
\end{align}

It should be noted that
the conventional
pulse shaping filters
(such as square root raised-cosine filters)
are not employed in practical OFDM systems
so as to preserve the orthogonality among subcarriers.
(The spectrum is usually controlled by windowing~\cite{hama2023achievable}.)
Thus,
the transmitted signal in~\eqref{eq:sc_ofdm} is strictly limited to the block time including CP,
i.e.,~$-N_\text{CP} T_0 \leq t \leq T_\text{block}$.
On the other hand,
its power spectrum does not localize in the frequency domain,
where
the mainlobe bandwidth is given by~\cite{hama2023achievable}
\begin{align}
  B
  &=
  \frac{N}{T_\text{block}}
  =
  \frac{N}{N T_0}
  =
  \frac{1}{T_0}.
  \label{eq:B_OFDM}
\end{align}

\subsection{Channel Model}

Throughout this work,
we consider doubly-selective channels,
where the signal is received through linear time variant~(LTV) channels~\cite{hlawatsch2011wireless}.
Let $h_p \in \mathbb{C}$, $\tau_p \in \mathbb{R}_{\geq 0}$, and~$\nu_p \in \mathbb{R}$ denote
the complex path gain, delay, and Doppler shift, respectively,
all corresponding to the $p$th path with~$p \in \left\{ 0, \cdots, P-1 \right\}$,
where $P$ represents the number of propagation paths
and $\mathbb{R}_{\geq 0}$ is a set of non-negative real numbers.
The channel delay-Doppler response is expressed as
\begin{align}
  h \left( \tau, \nu \right)
  &=
  \sum_{p=0}^{P-1} h_p \delta \left( \tau - \tau_p \right) \delta \left( \nu - \nu_p \right)
  ,
  \label{eq:DD_response}
\end{align}
where $\delta (\cdot)$ is the Dirac delta function.

As is often the case with delay-Doppler modulation context including OTFS~\cite{hadani2017orthogonal,raviteja2018practical},
we assume that the delay and Doppler shift
are integer multiples of
the sampling (Nyquist) interval~$T_0$
and the inverse of effective block time~$1/N T_0$, respectively.
Let $l_p \in \mathbb{Z}_{\geq 0}$ and $k_p \in \mathbb{Z}$
denote the integers representing delay and Doppler taps both corresponding to the $p$th path,
where $\mathbb{Z}_{\geq 0}$ is a set of non-negative integers.
Thus,
the delay time and Doppler frequency corresponding to the $p$th path are given by
\begin{align}
  \tau_p &=
  l_p T_0
  ,
  \quad
  \nu_p =
  \frac{k_p}{N T_0}
  .
  \label{eq:DD_integer}
\end{align}

In order to characterize doubly-selective channels,
we consider maximum values of delay and Doppler shift.
Let $l_{\text{max}} \in \mathbb{Z}_{\geq 0}$ and $k_{\text{max}} \in \mathbb{Z}_{\geq 0}$ denote
the maximum delay and Doppler taps, respectively.
They are defined as
\begin{align}
  l_\text{max}
  &\triangleq
  \max \left( l_0, \cdots, l_{P-1} \right),
  \label{eq:l_max}
  \\
  k_\text{max}
  &\triangleq
  \max \left( \left| k_0 \right|, \cdots, \left| k_{P-1} \right| \right).
  \label{eq:k_max}
\end{align}
Thus,
the maximum delay time and Doppler frequency for doubly-selective LTV channels,
denoted by~$\tau_\text{max} \in \mathbb{R}_{\geq 0}$ and~$\nu_\text{max} \in \mathbb{R}_{\geq 0}$,
are defined as
\begin{align}
  \tau_\text{max}
  &\triangleq
  l_\text{max}
  T_0
  ,
  \label{eq:tau_max}
  \\
  \nu_\text{max}
  &\triangleq
  \frac{k_\text{max}}{N T_0}
  ,
  \label{eq:nu_max}
\end{align}
respectively.

\subsection{Received Signal Representation}

The continuous-time received signal~$r (t)$ is expressed as~\cite{raviteja2018practical}
\begin{align}
  r \left( t \right)
  &=
  \int \int h \left( \tau, \nu \right) s \left( t - \tau \right) e^{j2\pi \nu \left( t-\tau \right)} d \tau d \nu
  + \eta \left( t \right)
  ,
  \label{eq:r_t}
\end{align}
where $\eta \left( t \right)$ denotes the additive white Gaussian noise~(AWGN) term.

By substituting~\eqref{eq:DD_response} and~\eqref{eq:DD_integer} into~\eqref{eq:r_t},
the $n$th discrete sample of the received symbol
after
CP removal
is given by
\begin{align}
  r_n
  &\triangleq
  r \left( n  T_0 \right)
  =
  \sum_{p=0}^{P-1} h_p e^{j2\pi \frac{k_p \left( n - l_p \right)}{N}} x_{\left[ n-l_p \right]_N} + \eta_n
  ,
  \nonumber\\&
  \qquad\qquad\qquad\qquad
  n \in \left\{ 0, \cdots, N-1 \right\}
  ,
  \label{eq:r_n}
\end{align}
where $\eta_n$ is the $n$th AWGN with zero mean and complex variance~$N_0$,
i.e.,~$\eta_n \sim \mathcal{CN} \left( 0, N_0 \right)$.
From~\eqref{eq:r_n},
the received signal-to-noise power ratio~(SNR) is defined as
\begin{align}
  \gamma_s
  &\triangleq
  \frac{E \left\{ \displaystyle \left| \sum_{p=0}^{P-1} h_p e^{j2\pi \frac{k_p \left( n - l_p \right)}{N}} x_{\left[ n-l_p \right]_N} \right|^2 \right\}}{E \left\{ \left| \eta_n \right|^2 \right\}}
  \nonumber\\&
  =
  \sum_{p=0}^{P-1} E \left\{ \left| h_p \right|^2 \right\}
  \frac{E_s}{N_0}
  ,
  \label{eq:snr}
\end{align}
where $E \left\{ \cdot \right\}$ represents the expectation operation.

In order to rewrite~\eqref{eq:r_n} in the vector form,
we define the matrix form of the LTV channel 
as~\cite{raviteja2018practical}
\begin{align}
  \mathbf{H}
  &=
  \sum_{p=0}^{P-1} h_p \boldsymbol\Pi^{l_p} \boldsymbol\Delta^{k_p},
  \label{eq:H}
\end{align}
where $\boldsymbol\Pi \in \{0, 1\}^{N \times N}$ is a permutation matrix representing to delay,
and $\boldsymbol\Delta \in \mathbb{C}^{N \times N}$ is a diagonal matrix representing to Doppler shift.
They are defined as
\begin{align}
  \boldsymbol\Pi &=
  \begin{bmatrix}
    0 & \cdots & 0 & 1 \\
    1 & \cdots & 0 & 0 \\
    \vdots & \ddots & \vdots & \vdots \\
    0 & \cdots & 1 & 0
  \end{bmatrix}
  ,\\
  \boldsymbol\Delta &=
  \text{diag}
  \left( e^{j2\pi \frac{0}{N}}, e^{j2\pi \frac{1}{N}}, \cdots, e^{j2\pi \frac{N-1}{N}} \right).
\end{align}
Then,
we may express
\begin{align}
  \mathbf{r}
  &=
  \mathbf{H} \mathbf{x} + \boldsymbol\eta
  ,
  \label{eq:r}
\end{align}
where $\mathbf{r} = \left( r_0, \cdots, r_{N-1} \right)^T \in \mathbb{C}^{N\times1}$
is the received symbol vector with its elements given by~\eqref{eq:r_n},
and $\boldsymbol\eta = \left( \eta_0, \cdots, \eta_{N-1} \right)^T \in \mathbb{C}^{N\times1}$
is the AWGN vector.


\section{Single-Carrier Delay-Doppler Domain Equalization}
\label{sec:sc_dde}

In this section,
we first introduce the discrete Zak transform~(DZT),
which converts time domain symbols into the DD domain.
Based on this transform,
we describe our proposed SC-DDE.

\subsection{Discrete Zak Transform}

\subsubsection{Definition and Properties}

The time domain symbols are mapped into the DD domain
by DZT~\cite{mohammed2021derivation}.
For the DZT operation,
we introduce the number of symbols in the delay domain~$L$ and that in the Doppler domain~$K$.
We note that $L$ and $K$ are chosen such that
their product matches the length of input sequence,
i.e.,~$LK = N$.
Let $\mathbf{u} = \left( u_0, \cdots, u_{N-1} \right)^T \in \mathbb{C}^{N \times 1}$ denote
the time domain symbol sequence with length~$N$.
We define the $l$th subvector of length~$K$ composed of the elements of~$\mathbf{u}$ with
equal sample spacing of~$L$ as
\begin{align}
  \mathbf{u}_{L,K}^{(l)}
  &\triangleq
  \left( u_l, u_{l+L}, \cdots, u_{l+(K-1)L} \right)^T \in \mathbb{C}^{K\times 1},
  \nonumber\\&
  \qquad\qquad\qquad
  l \in \left\{ 0, \cdots, L-1 \right\}.
  \label{eq:u_sub}
\end{align}
By using the $K$-point DFT,
the $(L,K)$-point DZT of~$\mathbf{u}$,
denoted by~$\mathcal{Z}_{(L,K)} \left\{ \mathbf{u} \right\} : \mathbb{C}^{LK\times 1} \to \mathbb{C}^{L\times K}$,
is defined as
\begin{align}
  &\mathcal{Z}_{(L,K)} \left\{ \mathbf{u} \right\}
  \triangleq
  \left[ \mathbf{F}_K \mathbf{u}_{L,K}^{(0)}, \mathbf{F}_K \mathbf{u}_{L,K}^{(1)}, \cdots, \mathbf{F}_K \mathbf{u}_{L,K}^{(L-1)} \right]^T
  \nonumber\\&
  =
  \begin{bmatrix}
    {\mathbf{u}_{L,K}^{(0)}}^T \\
    {\mathbf{u}_{L,K}^{(1)}}^T \\
    \vdots \\
    {\mathbf{u}_{L,K}^{(L-1)}}^T
  \end{bmatrix}
  \!
  \mathbf{F}_K
  =
  \underbrace{\begin{bmatrix}
    u_0 \!&\! u_{L} \!&\!\! \cdots \!\!&\! u_{(K-1)L} \\
    u_1 \!&\! u_{L+1} \!&\!\! \cdots \!\!&\! u_{(K-1)L+1} \\
    \vdots \!&\! \vdots \!&\!\! \ddots \!\!&\! \vdots \\
    u_{L-1} \!&\! u_{2L-1} \!&\!\! \cdots \!\!&\! u_{LK-1}
  \end{bmatrix}}_{\triangleq \mathbf{U} \in \mathbb{C}^{L\times K}}
  \!
  \mathbf{F}_K
  .
  \label{eq:Z_u}
\end{align}
Let $\mathbf{V} = \mathcal{Z}_{(L,K)} \left\{ \mathbf{u} \right\} \in \mathbb{C}^{L \times K}$ denote
the $(L,K)$-point DZT output for an input~$\mathbf{u} \in \mathbb{C}^{N \times 1}$.
Its $(l,k)$th element 
is expressed as~\cite{bolcskei1997discrete}
\begin{align}
  V_{l,k}
  &=
  \frac{1}{\sqrt{K}}
  \sum_{m=0}^{K-1} u_{l + m L}
  e^{-j2\pi \frac{k}{K} m}
  ,
  \nonumber\\&
  \quad
  l \in \left\{ 0, \cdots, L-1 \right\},
  \,\,
  k \in \left\{ 0, \cdots, K-1 \right\}
  .
  \label{eq:DZT}
\end{align}
In addition,
the $(L,K)$-point inverse DZT~(IDZT) of~$\mathbf{V}$,
denoted by~$\mathbf{u} = \mathcal{Z}^{-1}_{(L,K)} \left\{ \mathbf{V} \right\}: \mathbb{C}^{L \times K} \to \mathbb{C}^{LK\times 1}$,
is defined as
\begin{align}
  u_{l + k L}
  &=
  \frac{1}{\sqrt{K}}
  \sum_{m=0}^{K-1}
  V_{l,m}
  e^{j2\pi \frac{m}{K} k}
  ,
  \nonumber\\&
  \quad
  l \in \left\{ 0, \cdots, L-1 \right\},
  \,\,
  k \in \left\{ 0, \cdots, K-1 \right\}
  .
  \label{eq:IDZT}
\end{align}
Therefore, from the definition of~\eqref{eq:DZT},
the DZT is equivalent to the DFT performed for $K$ time domain points sampled at equal intervals
with distinct time offset values.

In the DZT operation,
we assume that the input sequence~$\mathbf{u}$ is periodic with its length~$N$.
Thus, from~\eqref{eq:DZT},
we have
\begin{align}
  \mathcal{Z}_{(L,K)} \left\{ \mathbf{u} \right\} [l,k+mK]
  &=
  \mathcal{Z}_{(L,K)} \left\{ \mathbf{u} \right\} [l,k]
  ,
  \label{eq:quasi_k}
  \\
  \mathcal{Z}_{(L,K)} \left\{ \mathbf{u} \right\} [l+mL,k]
  &=
  \mathcal{Z}_{(L,K)} \left\{ \mathbf{u} \right\} [l,k] e^{j2 \pi \frac{k}{K} m}
  ,
  \label{eq:quasi_l}
\end{align}
for~$m \in \mathbb{Z}$~\cite{bolcskei1997discrete}.

\subsubsection{Vectorized Discrete Zak Transform}

We define the vectorization of the DZT output matrix~$\mathbf{V}$ as
\begin{align}
  \mathbf{v}
  &\triangleq
  \text{vec} \left( \mathbf{V} \right)
  \nonumber\\&
  = \left( V_{0,0}, \cdots, V_{L-1,0}, V_{0,1}, \cdots, V_{LK-1,LK-1} \right)^T
  ,
  \label{eq:v}
\end{align}
where $\text{vec} \left( \cdot \right)$ is a vectorization of a matrix
such that
$\mathbf{a} = \text{vec} \left( \mathbf{A} \right) $
converts a matrix~$\mathbf{A}$ into a vector~$\mathbf{a}$
by stacking the columns of the matrix~$\mathbf{A}$ on top of one another.
Note that its $(l+kL)$th element is
equal to~$V_{l,k}$
given in~\eqref{eq:DZT}.
From the matrix form of the input vector~$\mathbf{u}$ defined as~$\mathbf{U}$ in~\eqref{eq:Z_u}, 
the vectorization of the DZT output of~\eqref{eq:v} can be expressed as
\begin{align}
  \mathbf{v}
  &=
  \text{vec} \left( \mathcal{Z}_{(L,K)} \left\{ \mathbf{u} \right\} \right)
  =
  \text{vec} \left( \mathbf{U} \mathbf{F}_K \right)
  =
  \text{vec} \left( \mathbf{I}_{L} \mathbf{U} \mathbf{F}_K \right)
  \nonumber\\&
  =
  \left( \mathbf{F}_K \otimes \mathbf{I}_L \right)
  \text{vec} \left( \mathbf{U} \right)
  =
  \left( \mathbf{F}_K \otimes \mathbf{I}_L \right)
  \mathbf{u}
  ,
  \label{eq:v}
\end{align}
since $\text{vec} \left( \mathbf{A} \mathbf{B} \mathbf{C} \right) = \left( \mathbf{C}^T \otimes \mathbf{A} \right) \text{vec} \left( \mathbf{B} \right)$
for any combination of matrices~$\mathbf{A} \in \mathbb{C}^{L \times M}$, $\mathbf{B} \in \mathbb{C}^{M \times N}$, and $\mathbf{C} \in \mathbb{C}^{N \times K}$,
where $\otimes$ denotes the Kronecker product.
Therefore,
the vectorization of the DZT output can be expressed by a linear operation~$\mathbf{F}_K \otimes \mathbf{I}_L$,
and we call this operation the {\it vectorized discrete Zak transform~(VDZT)}.
Accordingly,
we introduce the $(L,K)$-point VDZT matrix,
denoted by~$\mathbf{Z}_{L,K} \in \mathbb{C}^{LK \times LK}$,
that transforms the input vector of length~$N = LK$ into the DD domain of size~$L \times K$,
which is defined as
\begin{align}
  \mathbf{Z}_{L,K}
  &\triangleq
  \mathbf{F}_{K} \otimes \mathbf{I}_L
  .
  \label{eq:VDZT_matrix}
\end{align}
We note that
this VDZT matrix is unitary
since
$\mathbf{Z}_{L,K} \mathbf{Z}_{L,K}^\text{H} = \left( \mathbf{F}_K \mathbf{F}_K^\text{H} \right) \otimes \left( \mathbf{I}_L \mathbf{I}_L \right) = \mathbf{I}_{LK}$.
By using the VDZT matrix of~\eqref{eq:VDZT_matrix},
the VDZT operation of~\eqref{eq:v}
can be expressed as
\begin{align}
  \mathbf{v}
  &=
  \text{vec} \left( \mathcal{Z}_{(L,K)} \left\{ \mathbf{u} \right\} \right)
  =
  \mathbf{Z}_{L,K} \mathbf{u}
  .
  \label{eq:w_VDZT}
\end{align}

\begin{figure}[tb]
  \centering
   \includegraphics[width = \hsize, clip]{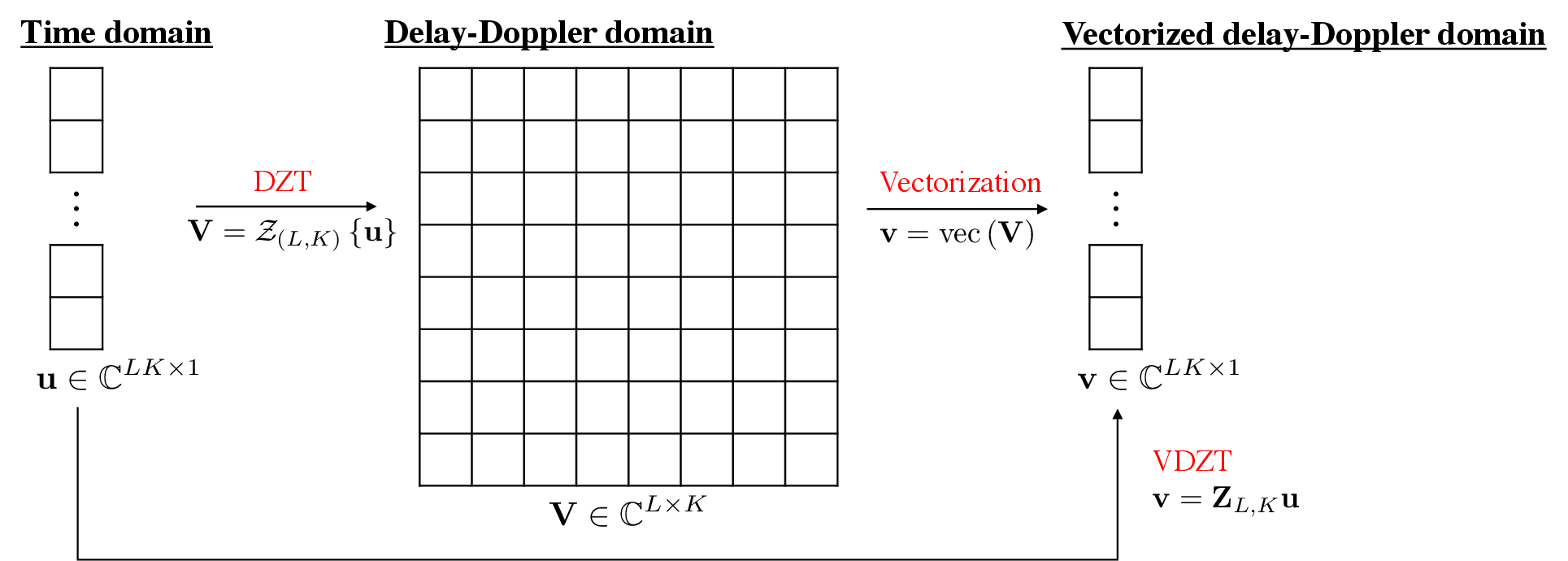}
   \caption{Symbol representation in the time, DD, and vectorized DD domain,
     where we set $N = 64$ and $(L, K) = (8, 8)$.}
  \label{dzt_vdzt}
\end{figure}

For reference,
Fig.~\ref{dzt_vdzt} illustrates the symbol representation
in the time and DD domain,
and we also depict the vectorization of the DZT output referred to as vectorized DD domain.
Here,
we consider the input symbol sequence with the length of~$N = LK = 64$,
and it is converted into the DD domain with~$(L, K) = (8, 8)$
by the $(8,8)$-point DZT operation.

\subsection{Delay-Doppler Domain Equalization}

\begin{figure}[tb]
  \centering
   \includegraphics[width = \hsize, clip]{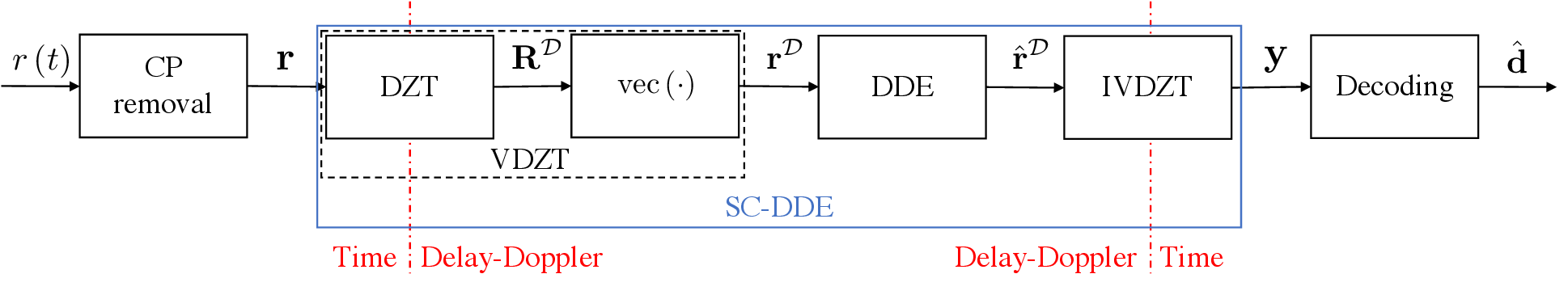}
   \caption{A block diagram of the SC-DDE receiver.}
  \label{receiver}
\end{figure}

We describe the system model of the SC-DDE receiver by using the DZT,
where its block diagram is shown in Fig.~\ref{receiver}.
In order to employ the equalization in the DD domain,
the received symbol is first mapped into the DD domain by the $(L,K)$-point DZT operation,
and the received symbol matrix in the DD domain is given by
\begin{align}
  \mathbf{R}^\mathcal{D}
  &=
  \mathcal{Z}_{(L,K)} \left\{ \mathbf{r} \right\}
  ,
  \label{eq:R_D}
\end{align}
where $\mathbf{r}$ is the received symbol vector in~\eqref{eq:r}.
In addition,
the $(l,k)$th received symbol in the DD domain is expressed as
\begin{align}
  R_{l,k}^\mathcal{D} 
  &=
  \mathbf{R}^\mathcal{D} [l,k]
  =
  \frac{1}{\sqrt{K}}
  \sum_{n=0}^{K-1} r_{l + n L}
  e^{-j2\pi \frac{k}{K} n}
  ,
  \nonumber\\&
  \qquad
  l \in \left\{ 0, \cdots, L-1 \right\},
  \,\,
  k \in \left\{ 0, \cdots, K-1 \right\}
  ,
  \label{eq:r_lk_DD}
\end{align}
where $r_n$ is the $n$th discretized received symbol given by~\eqref{eq:r_n}.
As a result,
delay and Doppler resolutions of the SC signal are given by~$T_0$ and~$1/ \left( NT_0 \right)$, respectively~\cite{lin2022orthogonal}.
Hence,
the $(l,k)$th received symbol in the DD domain~$R^\mathcal{D}_{l,k}$
is mapped onto the $(l,k)$th DD resource grid~$\Lambda^\mathcal{D}$
defined as~\cite{lin2022orthogonal}
\begin{align}
  &\Lambda^\mathcal{D}
  \!
  \triangleq
  \!
  \left\{
  \!\!
  \left(\! lT_0, \frac{k}{NT_0} \!\right)
  \!
  :
  l \! \in \! \{ 0, \! \cdots \! , \! L\!-\!1 \},
  \,
  k \! \in \! \{ 0, \! \cdots \! , \! K\!-\!1 \}
  \!
  \right\}
  .
  \label{eq:Lambda_DD}
\end{align}
{\it Remark:}
Since the delay and Doppler taps of the channel
may take~$0 \leq l_p \leq l_\text{max}$ and~$-k_\text{max} \leq k_p \leq k_\text{max}$,
the delay-Doppler grid in~\eqref{eq:Lambda_DD} must be larger than these ranges.
As a result,
the parameters~$L$ and~$K$ should be chosen
to meet the following condition:
\begin{align}
  \begin{cases}
  L > l_\text{max},
  \\
  K > 2k_\text{max}.
  \end{cases}
  \label{eq:LK_condition}
\end{align}

From~\eqref{eq:w_VDZT},
the vectorization of the received symbol matrix in the DD domain~$\mathbf{R}^\mathcal{D}$ in~\eqref{eq:R_D},
denoted by~$\mathbf{r}^\mathcal{D} \in \mathbb{C}^{N\times1}$,
can be expressed by the VDZT matrix as
\begin{align}
  \mathbf{r}^\mathcal{D}
  &\triangleq
  \text{vec} \left( \mathbf{R}^\mathcal{D} \right)
  \nonumber\\&
  =
  \left( R^\mathcal{D}_{0,0}, \cdots, R^\mathcal{D}_{L-1,0}, R^\mathcal{D}_{0,1}, \cdots, R^\mathcal{D}_{L-1,K-1} \right)^T
  \nonumber\\&
  =
  \mathbf{Z}_{L,K} \mathbf{r}
  .
\end{align}
SC-DDE is performed for the received symbol vector in the DD domain~$\mathbf{r}^\mathcal{D}$.
Thus,
we introduce the equivalent channel matrix observed in the DD domain,
denoted by~$\mathbf{H}^\mathcal{D} \in \mathbb{C}^{N \times N}$,
which is given from that in the time domain 
as~\cite{raviteja2018practical}
\begin{align}
  \mathbf{H}^\mathcal{D}
  &\triangleq
  \mathbf{Z}_{L,K} \mathbf{H} \mathbf{Z}_{L,K}^\text{H}
  =
  \left( \mathbf{F}_K \otimes \mathbf{I}_L \right) \mathbf{H} \left( \mathbf{F}_K^\text{H} \otimes \mathbf{I}_L \right)
  .
  \label{eq:H_DD}
\end{align}
It should be noted here that
unlike the conventional OFDM transmission without Doppler shifts,
delay-Doppler modulation,
such as OTFS and ODDM,
does not follow 
eigenmode transmission.
Therefore,
the equivalent channel matrix in the DD domain~$\mathbf{H}^\mathcal{D}$ in~\eqref{eq:H_DD}
does not become a diagonal matrix
even though the orthogonal pulse in the DD domain is employed~\cite{lin2022orthogonal}.
Consequently,
unlike one-tap equalization in OFDM and SC-FDE,
two-dimensional equalization is required in DDE similar to MIMO signal detection.

In this work,
we adopt linear equalization based on MMSE for DDE,
and its weight matrix~$\mathbf{W}^\mathcal{D} \in \mathbb{C}^{N \times N}$ is expressed as~\cite{nimr2018extended}
\begin{align}
  \mathbf{W}^\mathcal{D} &=
  \left( {\mathbf{H}^\mathcal{D}}^\text{H} \mathbf{H}^\mathcal{D} + \frac{1}{\gamma} \mathbf{I}_{N} \right)^{-1} {\mathbf{H}^\mathcal{D}}^\text{H}
  .
  \label{eq:W_DD}
\end{align}
After DDE,
the equalized received symbol vector in the DD domain is given by
\begin{align}
  \hat{\mathbf{r}}^\mathcal{D}
  &=
  \mathbf{W}^\mathcal{D} \mathbf{r}^\mathcal{D}
  .
  \label{eq:hat_r_DD}
\end{align}
Finally, after inverse VDZT~(IVDZT) postcoding,
the resulting estimated symbol vector~$\mathbf{y} \in \mathbb{C}^{N \times 1}$
is expressed by
\begin{align}
  \mathbf{y}
  &=
  \mathbf{Z}_{L,K}^\text{H} \hat{\mathbf{r}}^\mathcal{D}
  =
  \mathbf{Z}_{L,K}^\text{H} \mathbf{W}^\mathcal{D} \mathbf{Z}_{L,K} \mathbf{r}
  \nonumber\\&
  =
  \mathbf{Z}_{L,K}^\text{H} \mathbf{W}^\mathcal{D} \mathbf{H}^\mathcal{D} \mathbf{Z}_{L,K} \mathbf{x} + \mathbf{Z}_{L,K}^\text{H} \mathbf{W}^\mathcal{D} \mathbf{Z}_{L,K} \boldsymbol\eta
  .
  \label{eq:y}
\end{align}

\subsection{Conventional OTFS Modulation}
\label{sec:otfs}


\begin{figure*}[tb]
  \centering
   \includegraphics[width = 0.85\hsize, clip]{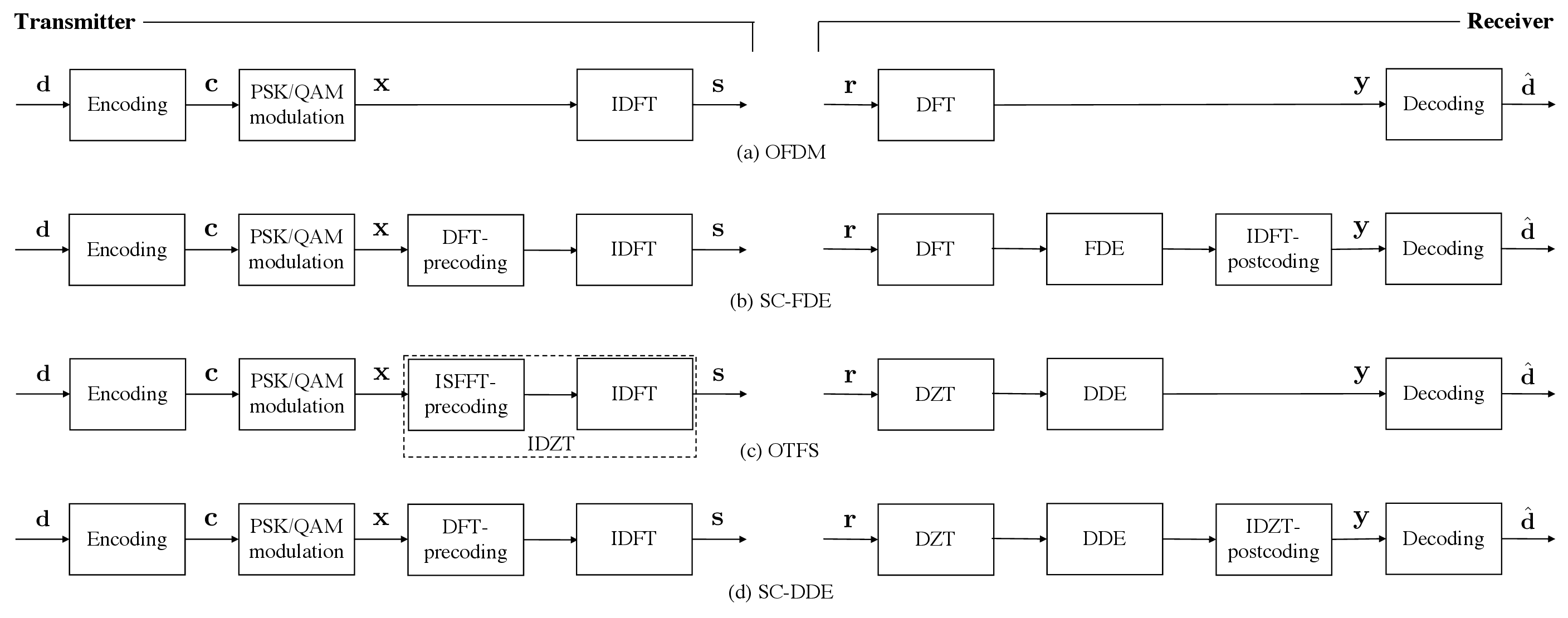}
   \caption{System block diagram comparison of (a)~OFDM, (b)~SC-FDE, (c)~OTFS, and (d)~SC-DDE,
   where the vectorization process is omitted.}
   \label{system_comparison}
\end{figure*}

For comparison,
we describe the conventional OTFS modulation
as delay-Doppler modulation scheme compensating doubly-selective channels.
Our proposed SC-DDE employing the $(L,K)$-point DZT corresponds to
OTFS modulation with $K$ multiple OFDM symbols each containing $L$ subcarriers.
Let $T_\text{MC} \triangleq T_\text{block} / K$ denote the symbol duration of each OFDM symbol.
In this case,
the subcarrier spacing is given by~$1/T_\text{MC} = K/T_\text{block}$
and its mainlobe bandwidth is expressed as~$B_\text{MC} = L/T_\text{MC} = N / T_\text{block} = 1/T_0$~\cite{hama2023achievable}.
Thus,
the resulting mainlobe bandwidth of OFDM and OTFS is equivalent to that of SC transmission in~\eqref{eq:B_OFDM},
and OFDM and OTFS carry $N = LK$ modulated symbols in each block.

Let $\mathbf{X}_\text{OTFS} \in \mathcal{X}^{L \times K}$ denote the modulated symbol matrix of OTFS in the DD domain.
Since OTFS modulation is identical to OFDM with the inverse symplectic finite Fourier transform~(ISFFT) precoding,
the continuous-time transmitted signal of OTFS is expressed as
\begin{align}
  s \left( t \right)
  &=
  \sum_{k=0}^{K-1}
  \sum_{l=0}^{L-1}
  \check{x}_{l,k}
  w_{[0,T_\text{MC}]} \left( t - kT_\text{MC} \right)
  e^{j 2\pi l (t-kT_\text{MC}) / T_\text{MC}}
  ,
  \label{eq:ofdm_signal}
\end{align}
where $\check{x}_{l,k}$ represents the $(l,k)$th ISFFT-precoded modulated symbol given by
\begin{align}
  \check{x}_{l,k}
  &=
  \frac{1}{\sqrt{LK}}
  \sum_{k'=0}^{K-1}
  \sum_{l'=0}^{L-1}
  \mathbf{X}_\text{OTFS} \left[ l',k' \right]
  e^{j 2\pi \left( \frac{k'k}{K} - \frac{l'l}{L} \right)}
  .
  \label{eq:mc_modulated_symbol}
\end{align}
In addition,
the transmitted symbol vectors of OTFS without oversampling,
denoted by~$\mathbf{s}_\text{OTFS} 
\in \mathbb{C}^{N\times 1}$,
is expressed by a vector form as~\cite{raviteja2018practical}
\begin{align}
  \mathbf{s}_\text{OTFS}
  &=
  \text{vec} \left( \mathbf{F}^\text{H}_L \left( \mathbf{F}_L \mathbf{X}_\text{OTFS} \mathbf{F}^\text{H}_K \right) \right)
  =
  \text{vec} \left( \mathbf{X}_\text{OTFS} \mathbf{F}^\text{H}_K \right)
  \nonumber\\&
  =
  \text{vec} \left( \mathcal{Z}_{L,K}^{-1} \left\{ \mathbf{X}_\text{OTFS} \right\} \right)
  =
  \mathbf{Z}^\text{H}_{L,K} \mathbf{x}_\text{OTFS}
  ,
  \label{eq:s_OTFS}
\end{align}
where $\mathbf{x}_\text{OTFS} \triangleq \text{vec} \left( \mathbf{X}_\text{OTFS} \right) \in \mathcal{X}^{LK \times 1}$. 
From~\eqref{eq:s_OTFS},
it is apparent that
OTFS modulation can be directly performed by the IDZT operation~\cite{lampel2022otfs}.


By replacing~$\mathbf{x}$ with~$\mathbf{s}_\text{OTFS}$ in~\eqref{eq:r},
the received symbol vector of OTFS is expressed as
\begin{align}
  \mathbf{r}_\text{OTFS} &= \mathbf{H} \mathbf{s}_\text{OTFS} + \boldsymbol\eta.
  \label{eq:r_OTFS}
\end{align}
The OTFS receiver performs the same signal processing as the SC-DDE receiver except for the postcoding.
Therefore,
the resulting received symbol vector of OTFS after DDE is expressed as
\begin{align}
  \mathbf{y}_\text{OTFS}
  &=
  \mathbf{W}^\mathcal{D} \mathbf{Z}_{L,K} \mathbf{r}_\text{OTFS}
  \nonumber\\&
  =
  \mathbf{W}^\mathcal{D} \mathbf{Z}_{L,K} \left( \mathbf{H} \mathbf{Z}^\text{H}_{L,K} \mathbf{x}_\text{OTFS} + \boldsymbol\eta \right)
  \nonumber\\&
  =
  \mathbf{W}^\mathcal{D} \mathbf{H}^\mathcal{D} \mathbf{x}_\text{OTFS} + \mathbf{W}^\mathcal{D} \mathbf{Z}_{L,K} \boldsymbol\eta
  .
  \label{eq:y_OTFS}
\end{align}

\begin{table*}[tb]
  \begin{center}
    \caption{Computational complexity order of OFDM, SC-FDE, OTFS, and SC-DDE transceivers ($N = LK$).}
    \begin{tabular}{c | c | c c c c}
      & {\bf Transmitter} & \multicolumn{4}{|c}{\bf Receiver} \\
      System & \multicolumn{1}{c|}{Modulation} & Domain transform & (Linear) Equalization & MMSE weight calc. & Postcoding
      \\ \hline \hline
      {\bf OFDM} & $N$-point IDFT & $N$-point DFT & - & - & - \\
      {\bf SC-FDE} & $N$-point DFT\&IDFT & $N$-point DFT & $\mathcal{O}\left( N \right)$ & $\mathcal{O}\left( N \right)$ & $N$-point IDFT \\
      {\bf OTFS} & $(L,K)$-point IDZT & $(L,K)$-point DZT & $\mathcal{O}\left( N \log N + N \right)$ & $\mathcal{O}\left( N^3 \right)$ & - \\
      {\bf SC-DDE} & $N$-point DFT\&IDFT & $(L,K)$-point DZT & $\mathcal{O}\left( N \log N + N \right)$ & $\mathcal{O}\left( N^3 \right)$ & $(L,K)$-point IDZT
    \end{tabular}
    \label{table:complexity}
  \end{center}
  \vspace{-3.0mm}
\end{table*}

\begin{table}[tb]
  \begin{center}
    \caption{Computational complexity order of DFT and DZT.}
    \begin{tabular}{c c c}
      Operation & Complexity w/o FFT & Complexity w/ FFT
      \\ \hline \hline
      $N$-point DFT/IDFT & $\mathcal{O} \left( N^2 \right)$ & $\mathcal{O} \left( N \log N \right)$ \\
      $(L,K)$-point DZT/IDZT & $\mathcal{O} \left( L K^2 \right)$ & $\mathcal{O} \left( L K \log K \right)$
    \end{tabular}
    \label{table:complexity_dft_dzt}
  \end{center}
  \vspace{-3.0mm}
\end{table}

\subsection{Complexity}

The main drawback of SC-DDE system is its high computational complexity
for weight matrix calculation in~\eqref{eq:W_DD} and equalization in~\eqref{eq:hat_r_DD}.
In this work,
we define the computational complexity as
the number of complex multiplications,
and we evaluate the complexity order of SC-DDE transmitter and receiver.
For comparison,
we introduce OTFS modulation described in Section~\ref{sec:otfs},
which corresponds to the ISFFT-precoded $K$ multiple OFDM symbols each containing $L$ subcarriers.
In addition,
we also make comparison with practical OFDM and SC-FDE systems,
which are adopted in the downlink and uplink of modern cellular networks, respectively.
The block diagrams of these schemes are shown in Fig.~\ref{system_comparison}.
In this subsection,
we omit the vectorization operation
since it does not involve arithmetic operations, 
i.e., DZT and VDZT are not distinguished.

We first discuss the transmitter complexity.
From Fig.~\ref{system_comparison},
we observe that our SC-DDE transmitter is identical to that of SC-FDE,
which requires DFT-precoding.
Therefore,
its complexity is higher than that of OFDM.

We next compare the receiver complexity
assuming that 
linear equalization based on MMSE is employed for all schemes.
In SC-DDE and OTFS,
the received symbol sequence is first mapped into the DD domain by the DZT,  
whereas it is mapped into the frequency domain by the DFT in SC-FDE and OFDM.
Since delay-Doppler modulation is not based on eigenmode transmission,
OTFS receiver requires 2D equalization similar to SC-DDE.
Therefore,
in SC-DDE and OTFS systems,
the complexity order for equalization in~\eqref{eq:hat_r_DD}
and that for MMSE matrix calculation in~\eqref{eq:W_DD}
are $\mathcal{O} \left( N \log N + N \right)$
and $\mathcal{O} \left( N^3 \right)$,
respectively~\cite{surabhi2019low}.
In contrast,
SC-FDE can employ one-tap equalization
based on the circulant structure of the channel matrix
without Doppler shifts.
Therefore,
its complexity order in the SC-FDE system is only $\mathcal{O} \left( N \right)$ for equalization
and $\mathcal{O} \left( N \right)$ for MMSE weight calculation.
Furthermore,
in the OFDM system,
equalization is not necessary
since the likelihood information can be directly calculated from the received symbol sequence.
Also,
the receiver of SC-DDE and SC-FDE requires postcoding after equalization
by IDZT and IDFT, respectively,
since the domain of modulated symbols is different from that where equalization process takes place.
As a result,
the receiver complexity of the proposed SC-DDE is higher than that of OTFS by the IDZT postcoding,
and significantly higher than SC-FDE and OFDM due to 2D equalization.

Based on the above observation,
we summarize the complexity order of the transmitter and receiver in
OFDM, SC-FDE, OTFS, and SC-DDE systems in Table~\ref{table:complexity},
where that of DFT and DZT operations is listed in Table~\ref{table:complexity_dft_dzt}.
We note that
the complexity of $(L,K)$-point DZT is $L$ times that of $K$-point DFT,
which is obvious from~\eqref{eq:Z_u}. 
As a result,
it is apparent from Table~\ref{table:complexity} that
the transmitter complexity of SC-DDE is equivalent to that of SC-FDE,
but the SC-DDE receiver requires considerable additional complexity compared to the conventional FDE.
In conclusion,
our proposed SC-DDE is a suitable approach especially for uplink communications
similar to the conventional SC-FDE.

\section{Channel Estimation Designed for SC-DDE}
\label{sec:channel_estimation}

In this section,
we introduce the channel estimation designed for SC-DDE.
In the conventional channel estimation for SC transmission in cellular networks,
the pilot symbols are transmitted in the frequency domain
so as to estimate the frequency impulse response required for FDE.
However,
this approach is not applicable to our proposed SC-DDE
since DDE requires the channel delay Doppler response.
Hence,
we consider the embedded pilot-aided channel estimation proposed for OTFS modulation~\cite{raviteja2019embedded},
where the data symbols with PSK/QAM signaling are transmitted simultaneously
with the embedded pilot symbols in the DD domain.
In the case of OTFS modulation,
they are generated in the same DD domain.
However,
in our proposed SC-DDE system,
the pilot symbol should be transmitted in the DD domain
so as to estimate the channel delay Doppler response,
whereas the data symbols are modulated by PSK/QAM signaling in the time domain.
Therefore,
the embedded pilot-aided channel estimation must be designed for SC-DDE
so that the data and pilot symbols generated in the different domains
do not interfere with each other. 

To this end,
we derive the delay-Doppler domain representation of SC transmission.
By utilizing this property,
we introduce the embedded pilot-aided channel estimation designed for SC-DDE.

\subsection{Delay-Doppler Domain Input-Output Relation for Single-Carrier Transmission}

Since the channel estimation for SC-DDE is performed in the DD domain,
the pilot symbol should also be embedded in the same domain.
Thus,
in this subsection,
we derive the relation between
$\mathbf{R}^\mathcal{D}$ given by~\eqref{eq:r_lk_DD}
and the DD domain symbol matrix of the modulated symbol vector denoted by~$\mathbf{X}^\mathcal{D} \in \mathbb{C}^{L \times K}$,
which is expressed by the DZT operation as
\begin{align}
  \mathbf{X}^\mathcal{D}
  =
  \mathcal{Z}_{(L,K)} \left\{ \mathbf{x} \right\}
  .
  \label{eq:X_DD_lk}
\end{align}

From~\eqref{eq:r_n},
the $n$th received symbol is rewritten as
\begin{align}
  r_n
  &=
  \sum_{p=0}^{P-1} h_p \tilde{r}_{p,n} + \eta_n
  ,
  \label{eq:r_n_2}
\end{align}
where $\tilde{r}_{p,n}$ represents the $n$th received symbol
corresponding to the $p$th path defined as
\begin{align}
  \tilde{r}_{p,n}
  &\triangleq
  e^{j2 \pi \frac{k_p \left( n-l_p \right)}{N}} x_{\left[ n-l_p \right]_N}
  .
  \label{eq:r_p_n}
\end{align}
Thus,
by introducing the received symbol vector corresponding to the $p$th path,
denoted by~$\tilde{\mathbf{r}}_{p} = \left( \tilde{r}_{p,0}, \cdots, \tilde{r}_{p,N-1} \right)^T \in \mathbb{C}^{N \times 1}$,
the received symbol vector of~\eqref{eq:r} can be rewritten by
\begin{align}
  \mathbf{r}
  &=
  \sum_{p=0}^{P-1} h_p \tilde{\mathbf{r}}_{p} + \boldsymbol\eta
  .
  \label{eq:r_2}
\end{align}
Since the DZT is a linear transform~\cite{bolcskei1997discrete},
the DZT of the received symbol vector~$\mathbf{r}$
is expressed by using~\eqref{eq:r_2} as
\begin{align}
  \mathbf{R}^\mathcal{D}
  &=
  \mathcal{Z}_{(L,K)} \left\{ \mathbf{r} \right\}
  \nonumber\\&
  =
  \sum_{p=0}^{P-1} h_p
  \mathcal{Z}_{(L,K)} \left\{ \tilde{\mathbf{r}}_p \right\}
  +
  \mathcal{Z}_{(L,K)} \left\{ \boldsymbol\eta \right\}
  \nonumber\\&
  =
  \sum_{p=0}^{P-1} h_p
  \mathcal{Z}_{(L,K)} \left\{ \tilde{\mathbf{r}}_p \right\}
  +
  \boldsymbol\Omega
  ,
  \label{eq:DD_IO_1}
\end{align}
where the noise term in the DD domain~$\boldsymbol\Omega \triangleq \mathcal{Z}_{(L,K)} \left\{ \boldsymbol\eta \right\}$
is also AWGN with~$\boldsymbol\Omega[l,k] \sim \mathcal{CN} \left( 0, N_0 \right)$,
since the DZT is a unitary transform~\cite{bolcskei1997discrete}.
From {\it quasi}-periodic property in~\eqref{eq:quasi_k} and~\eqref{eq:quasi_l}
and the relationship of~$\mathcal{Z}_{(L,K)} \left\{ \mathbf{x} e^{j2\pi m n / N} \right\} [l,k] = e^{j2\pi m l / N} \mathcal{Z}_{(L,K)} \left\{ \mathbf{x} \right\} [l, k-m]$~\cite{bolcskei1997discrete},
the DZT of the received symbol vector corresponding to the $p$th path~$\tilde{\mathbf{r}}_p$ in~\eqref{eq:DD_IO_1} is expressed from~\eqref{eq:r_p_n} as
\begin{align}
  &\mathcal{Z}_{(L,K)} \left\{ \tilde{\mathbf{r}}_p \right\} [l,k]
  =
  e^{j2\pi k_p \frac{[ l - l_p ]_L}{N}}
  \mathcal{Z}_{(L,K)} \left\{ \mathbf{x} \right\} [l-l_p, k-k_p]
  \nonumber\\&
  =
  e^{j2\pi k_p \frac{[ l - l_p ]_L}{N}}
  \begin{cases}
    X_{[ l - l_p ]_L, [ k-k_p ]_K}^\mathcal{D}
    ,
    & l \geq l_p, \\
    X_{[ l - l_p ]_L, [ k-k_p ]_K}^\mathcal{D}
    e^{-j 2 \pi \frac{k}{K}}
    ,
    & l < l_p,
  \end{cases}
  \nonumber\\&
  \qquad\qquad
  l \in \left\{ 0, \cdots, L-1 \right\},
  \,\,
  k \in \left\{ 0, \cdots, K-1 \right\}
  .
  \label{eq:Z_r_p}
\end{align}

Therefore,
from~\eqref{eq:DD_IO_1} and~\eqref{eq:Z_r_p},
the resulting DD domain input-output relation of SC transmission
is expressed as
\begin{align}
  R_{l,k}^\mathcal{D}
  \!
  &=
  \!
  \begin{cases}
    \displaystyle\sum_{p=0}^{P-1} h_p e^{j2\pi k_p \frac{[ l - l_p ]_L}{N}}
    X_{[ l - l_p ]_L, [ k-k_p ]_K}^\mathcal{D}
    \! + \Omega_{l,k}
    ,
    \\ \qquad\qquad\qquad\qquad\qquad\qquad\qquad\qquad\qquad
    l \geq l_p, \\
    \displaystyle\sum_{p=0}^{P-1} h_p e^{j2\pi \left( k_p \frac{[ l - l_p ]_L}{N} - \frac{k}{K} \right)}
    X_{[ l - l_p ]_L, [ k-k_p ]_K}^\mathcal{D}
    \! + \Omega_{l,k}
    ,
    \\ \qquad\qquad\qquad\qquad\qquad\qquad\qquad\qquad\qquad
    l < l_p,
  \end{cases}
  \nonumber\\&
  \qquad\quad
  l \in \left\{ 0, \cdots, L-1 \right\},
  \,\,
  k \in \left\{ 0, \cdots, K-1 \right\}
  .
  \label{eq:DD_IO}
\end{align}

As a result,
the $(l,k)$th received symbol in the DD domain can be expressed by
a linear combination of
$(l-l_p, k-k_p)$th received symbols in the same domain
for a given pair of delay tap~$l_p$ and Doppler tap~$k_p$
corresponding to the $p$th path with~$p \in \left\{ 0, \cdots, P-1 \right\}$.
By utilizing this property,
we propose a channel estimation scheme for SC-DDE in the next subsection.

\subsection{Embedded Pilot-Aided Channel Estimation for SC-DDE}

The main objective of the channel estimation in the DD domain is
to estimate the $p$th complex channel gain~$h_p$
corresponding to the $l_p$ delay and $k_p$ Doppler taps
from~\eqref{eq:DD_IO}.
To this end,
we introduce an embedded pilot-aided channel estimation scheme designed for SC-DDE
by extending the approach proposed for OTFS in~\cite{raviteja2019embedded} to SC transmission.

Let $\mathbf{x}_\text{CE} = \left( x_{\text{CE}, 0}, \cdots, x_{\text{CE}, N-1} \right)^T \in \mathbb{C}^{N\times1}$
denote the time domain transmitted symbol vector of block length~$N$
with embedded pilot symbol for channel estimation of SC-DDE.
Since the pilot symbol is embedded in the data symbols,
it is expressed as
\begin{align}
  \mathbf{x}_\text{CE}
  &=
  \mathbf{x}_\text{data} + \mathbf{x}_\text{pilot}
  ,
  \label{eq:x_CE}
\end{align}
where $\mathbf{x}_\text{data} \in \left( \mathcal{X} \cup \left\{ 0 \right\} \right)^{N \times 1}$ and $\mathbf{x}_\text{pilot} \in \mathbb{C}^{N\times1}$ represent
the time domain vectors corresponding to data and pilot, respectively.
As in the conventional embedded pilot-aided channel estimation for OTFS,
we consider that one pilot symbol in the DD domain,
denoted by~$\psi^\mathcal{D} \in \mathbb{C}$,
is transmitted in each block~\cite{raviteja2019embedded},
and its power is given by~$E_\text{pilot}$
(i.e.,~$\left| \psi^\mathcal{D} \right|^2 = E_\text{pilot}$).
As a result,
the pilot symbol matrix in the DD domain is expressed as
\begin{align}
  \mathbf{X}^\mathcal{D}_{\text{pilot}} [l,k]
  &\triangleq
  \mathcal{Z}_{(L,K)} \left\{ \mathbf{x}_\text{pilot} \right\} [l,k]
  \nonumber\\&
  =
  \begin{cases}
    \psi^\mathcal{D}, & l = l_\text{pilot}, k = k_\text{pilot},\\
    0, & \text{otherwise},
  \end{cases}
\end{align}
where the pilot symbol~$\psi^\mathcal{D}$ is arranged on the $(l_\text{pilot}, k_\text{pilot})$th DD grid.
Thus,
by the IDZT operation,
the pilot symbol vector in the time domain is expressed as
\begin{align}
  \mathbf{x}_\text{pilot}
  =
  \mathcal{Z}_{(L,K)}^{-1}
  \left\{ \mathbf{X}^\mathcal{D}_{\text{pilot}} \right\}
  .
\end{align}

In order to distinguish between pilot and data symbols,
they must be located such that the receiver can separately extract them
from the received symbols in the DD domain.
Let $\mathbf{X}_\text{CE}^{\mathcal{D}} \triangleq \mathcal{Z}_{(L,K)} \left\{ \mathbf{x}_\text{CE} \right\} \in \mathbb{C}^{L \times K}$ denote
the DZT of the transmitted symbol vector~$\mathbf{x}_\text{CE}$.
From~\eqref{eq:DD_IO},
in order to avoid the interference among pilot and data symbols in the DD domain,
guard symbols (i.e., null) is inserted in advance in the DD grid where the pilot signal is received.
Thus,
guard symbols must be inserted on the $(l,k)$th DD grid satisfying
\begin{align}
  &\mathbf{X}^\mathcal{D}_\text{CE} \left[ l, k \right]
  = \mathbf{X}^\mathcal{D}_\text{CE} \left[ \left[l' \right]_L, \left[k' \right]_K \right]
  =
  0,
  \nonumber\\
  \begin{split}
    \quad
    \text{for}~
    \begin{cases}
      l' \neq l_\text{pilot},
      l_\text{pilot} - l_\text{max} \leq l' \leq l_\text{pilot} + l_\text{max},
      \\
      k' \neq k_\text{pilot},
      k_\text{pilot} - 2k_\text{max} \leq k' \leq k_\text{pilot} + 2k_\text{max},
    \end{cases}
  \end{split}
  \nonumber\\&
  \qquad\qquad
  l \in \left\{ \left[ l' \right]_L \middle| l' \in \mathbb{Z} \right\},
  \,\,
  k \in \left\{ \left[ k' \right]_K \middle| k' \in \mathbb{Z} \right\}.
  \label{eq:pilot_condition_DD}
\end{align}
This approach follows the same principle as that in the conventional OTFS
with embedded pilot-aided channel estimation~\cite{raviteja2019embedded}.

In addition to the condition in~\eqref{eq:pilot_condition_DD},
in the case of SC transmission,
modulated data symbols are spread out in the DD domain
since they are generated in the time domain.
Therefore,
SC signal generated in the time domain must satisfy
the condition in the DD domain given by~\eqref{eq:pilot_condition_DD}.
According to~\eqref{eq:u_sub},
we define the $l$th subvector of length~$K$ composed of the elements of~$\mathbf{x}_\text{CE}$ with equal sample spacing of~$L$ as
\begin{align}
  \mathbf{x}_{\text{CE}}^{(l)}
  &\triangleq
  \left( \mathbf{x}_\text{CE} [l], \mathbf{x}_{\text{CE}} [l+L], \cdots, \mathbf{x}_\text{CE} [l\!+\!(K\!-\!1)L] \right)^T \! \in \mathbb{C}^{K \times 1},
  \nonumber\\&
  \qquad\qquad\qquad
  l \in \left\{ 0, \cdots, L-1 \right\}
  .
  \label{eq:x_CE_sub}
\end{align}
Based on~\eqref{eq:Z_u},
the $(l,k)$th DZT output of the transmitted symbol vector~$\mathbf{x}_\text{CE}$
is expressed using~\eqref{eq:x_CE_sub} as
\begin{align}
  \mathbf{X}_{\text{CE}}^{\mathcal{D}} [l, k] 
  &=
  {\mathbf{x}_\text{CE}^{(l)}}^T
  \mathbf{f}_K^{(k)}
  ,
  \label{eq:DZT_independent}
\end{align}
where $\mathbf{f}_K^{(k)}$ represents the $k$th column vector of the DFT matrix with size~$K$.
Therefore,
it should be noted from~\eqref{eq:DZT_independent} that
DD domain symbols with the delay tap~$l \in \{ 0, \cdots, L-1 \}$
is determined only by the time domain symbols
corresponding to the $l$th subvector~$\mathbf{x}_\text{CE}^{(l)}$,
and independent of the other time domain symbols~$\mathbf{x}_\text{CE}^{(l')}$ with~$l' \neq l$.
On the other hand,
from~\eqref{eq:DZT_independent},
the time domain symbols corresponding to the $l$th subvector~$\mathbf{x}_\text{CE}^{(l)}$ are spread out in the Doppler domain.
Therefore, in the case of SC transmission,
the data symbols cannot be placed in the same delay tap as the pilot and guard symbols.
As a result,
the $l$th subvector of
the time domain data symbol vector,
denoted by $\mathbf{x}_{\text{data}}^{(l)} \triangleq \left( \mathbf{x}_\text{data} [l], \mathbf{x}_{\text{data}} [l+L], \cdots, \mathbf{x}_\text{data} [l\!+\!(K\!-\!1)L] \right)^T \! \in \mathbb{C}^{K \times 1}$,
should satisfy
\begin{align}
  &{\mathbf{x}_\text{data}^{(l)}}
  = {\mathbf{x}_\text{data}^{( [ l' ]_L)}}
  =
  \mathbf{0}_K,
  \quad
  \text{for }\,\,
  l_\text{pilot} - l_\text{max} \leq l' \leq l_\text{pilot} + l_\text{max},
  \nonumber\\&
  \qquad\qquad\qquad\qquad\qquad\qquad\qquad
  l \in \left\{ \left[ l' \right]_L \middle| l' \in \mathbb{Z} \right\}
  ,
  \label{eq:x_data}
\end{align}
where $\mathbf{0}_K$ represents the zero vector of length~$K$.
We note that modulated symbols can be transmitted
by the subvectors of the data symbol vector 
with their delay taps outside this condition.

\begin{figure*}[tb]
  \centering
   \includegraphics[width = 0.75\hsize, clip]{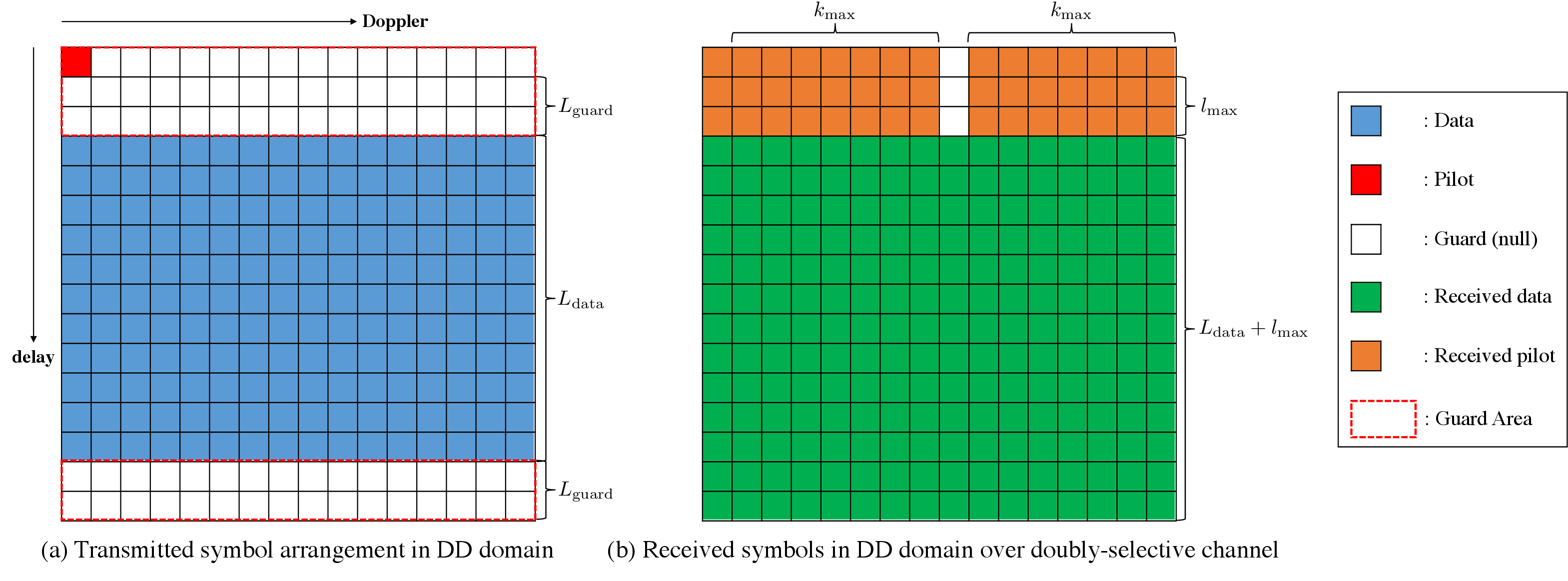}
   \caption{Pilot symbol configuration in the DD domain for SC-DDE,
       where $(L,K) = (16,16)$, $L_\text{guard} = 2$, and $(l_\text{max}, k_\text{max}) = (2, 7)$.}
  \label{pilot}
\end{figure*}

Motivated by the above observation,
we introduce the number of guard delay taps~$L_\text{guard} \in \mathbb{Z}_{\geq0}$,
and all data symbols are mapped $L_\text{guard}$ delay taps away from the pilot symbol.
It is designed for the maximum delay tap~$l_\text{max}$ so as to satisfy
\begin{align}
  L_\text{guard}
  &\geq
  l_\text{max}
  .
  \label{eq:L_guard}
\end{align}
Therefore,
from the conditions of~\eqref{eq:pilot_condition_DD} and~\eqref{eq:x_data}, 
the $(l,k)$th DD domain transmitted symbol of SC-DDE
with embedded pilot symbol should be arranged as
\begin{align}
  &\mathbf{X}^\mathcal{D}_\text{CE} \left[ l, k \right]
  = \mathbf{X}^\mathcal{D}_\text{CE} \left[ \left[l'\right]_L, k \right]
  =
  \nonumber\\& \quad
  \begin{cases}
    \psi^\mathcal{D}, & l' = l_\text{pilot}, k = k_\text{pilot}, \\
    0, & l_\text{pilot} - L_\text{guard} \leq l' \leq l_\text{pilot} + L_\text{guard},  \\
    {\mathbf{x}_\text{data}^{\left( l \right)}}^T \mathbf{f}_K^{(k)}, & \text{otherwise},
  \end{cases}
  \nonumber\\&
  \qquad\quad
  l \in \left\{ \left[ l' \right]_L \middle| l' \in \mathbb{Z} \right\},
  \,\,
  k \in \left\{ 0, \! \cdots \!, K\!-\!1 \right\}.
  \label{eq:DD_pilot}
\end{align}
Hence,
the number of modulated data symbols in each block is given by
\begin{align}
  N_\text{data}
  &=
  \left( L - 2L_\text{guard} -1 \right) K
  \triangleq
  L_\text{data} K
  ,
  \label{eq:N_data}
\end{align}
where $L_\text{data} \triangleq L - 2L_\text{guard} -1$ is the number of data symbols in the delay domain.
As a result,
the rate loss caused by embedded pilot symbol for SC-DDE is given by
\begin{align}
  \frac{N_\text{data}}{N}
  &=
  \frac{L_\text{data} K}{LK}
  =
  \frac{L - 2L_\text{guard} -1}{L}
  =
  1-
  \frac{2L_\text{guard} +1}{L}
  .
  \label{eq:rate_loss_pilot}
\end{align}

Fig.~\ref{pilot} illustrates the DD domain symbol configuration of SC-DDE with embedded pilot symbol
for an example of~$(L,K) = (16,16)$ and~$L_\text{guard} = 2$.
Fig.~\ref{pilot}(a) illustrates the transmitted symbol configuration.
Since the pilot symbol is located on the $(0,0)$th DD grid,
data symbols are mapped in the $(l, k)$th element in the DD domain
with~$l \in \{ L_\text{guard} + 1, \cdots, L-1-L_\text{guard} \}$ and~$k \in \{ 0, \cdots, K-1 \}$
according to the condition of~\eqref{eq:DD_pilot}.
In addition,
when the channel is characterized by~$(l_\text{max}, k_\text{max}) = (2, 7)$,
the received symbol matrix in the DD domain~$\mathbf{R}^\mathcal{D}$ are illustrated as Fig.~\ref{pilot}(b).
As a result,
pilot and data symbols do not interfere with each other
since \eqref{eq:L_guard} is satisfied.
Furthermore,
since $\left( l_\text{pilot}, k_\text{pilot} \right) = (0,0)$,
the $p$th estimated complex channel gain corresponding to $l_p$ delay tap and $k_p$ Doppler tap,
denoted by~$\hat{h}_p$,
is expressed as
\begin{align}
  \hat{h}_p
  &=
  \frac{R_{l_p, [k_p]_{K}}^\mathcal{D}}{\psi^\mathcal{D}}
  =
  h_p
  +
  \frac{\Omega_{l_p, [k_p]_K}}{\psi^\mathcal{D}}
  .
  \label{eq:hat_h}
\end{align}
Therefore,
the received SNR of the pilot symbol corresponding to the $p$th path is defined as
\begin{align}
  \gamma_{\text{pilot},p}
  &\triangleq
  E \left\{ \left| h_p \right|^2 \right\}
  \frac{E_\text{pilot}}{N_0}.
  \label{eq:SNR_p}
\end{align}

\subsection{PAPR Property With Embedded Pilot Symbol}

We now consider the time domain representation of the transmitted signal
in SC-DDE systems with embedded pilot symbol.
From the IDZT operation in~\eqref{eq:IDZT},
the $(l+kL)$th time domain pilot symbol is expressed as
\begin{align}
  \mathbf{x}_{\text{pilot}}[l+kL]  
  &=
  \begin{cases}
    \displaystyle\frac{1}{\sqrt{K}} \psi^\mathcal{D} e^{j2\pi k_\text{pilot} k}
    ,
    & l = l_\text{pilot},\\
    0
    ,
    & \text{otherwise},
  \end{cases}
  \nonumber\\&
  \quad\quad\quad\quad\quad\quad
  k \in \{ 0, \cdots, K-1 \}
  .
  \label{eq:pilot_time}
\end{align}
Hence,
by substituting~\eqref{eq:x_data} and~\eqref{eq:pilot_time} into~\eqref{eq:x_CE},
the $(l+kL)$th transmitted symbol vector of SC-DDE system with embedded pilot symbol
is expressed as
\begin{align}
  &\mathbf{x}_\text{CE} \left[ l + k L \right] 
  = \mathbf{x}_\text{CE} \left[ \left[ l' \right]_L + k L \right]
  =
  \nonumber\\&
  \begin{cases}
    \displaystyle\frac{1}{\sqrt{K}} \psi^\mathcal{D} e^{j2\pi k_\text{pilot} k}
    ,
    & l' = l_\text{pilot},\\
    \displaystyle0
    ,
    & l_\text{pilot} - L_\text{guard} \leq l' \leq l_\text{pilot} + L_\text{guard},\\
    \displaystyle \mathbf{x}[l+kL] 
    ,
    & \text{otherwise},
  \end{cases}
  \nonumber\\&
  \qquad\qquad
  l \in \left\{ \left[ l' \right]_L \middle| l' \in \mathbb{Z} \right\},
  \,\,
  k \in \left\{ 0, \cdots, K-1 \right\} 
  .
  \label{eq:SC_pilot}
\end{align}
It is important to note from~\eqref{eq:pilot_time} 
that
the time domain pilot symbol has the constant envelope of~$\left| \frac{1}{\sqrt{K}} \psi^\mathcal{D} e^{j2\pi k_\text{pilot} k} \right| = \left| \psi^\mathcal{D} \right| / \sqrt{K} = \sqrt{E_\text{pilot} / K}$ regardless of~$k \in \{ 0, \cdots, K-1 \}$.
Therefore,
when we set the pilot symbol power as~$E_\text{pilot} \leq K E_s$,
the embedded pilot symbol for channel estimation does not affect the peak power property of the SC transmission
given in~\eqref{eq:SC_pilot}.

For simplicity of description, in what follows,
we focus on the case with $\left( l_\text{pilot}, k_\text{pilot} \right) = \left( 0,0 \right)$,
i.e., the pilot symbol is arranged on the $(0,0)$th DD grid.
For a given number of guard symbols~$L_\text{guard}$,
from~\eqref{eq:SC_pilot},
the resulting transmitted symbol vector~$\mathbf{x}_\text{CE}$ is expressed as
\begin{align}
  \mathbf{x}_\text{CE}
  \!
  &=
  \!
  (
  \underbrace{\psi^\mathcal{D}/\sqrt{K}, \underbrace{0, \cdots, 0}_{L_\text{guard}}, \underbrace{x_{0}, \cdots, x_{L_\text{data}-1}}_{L_\text{data}}, \underbrace{0, \cdots, 0}_{L_\text{guard}}}_{L},
  \nonumber\\&
  \quad\,\,
  \underbrace{\psi^\mathcal{D}/\sqrt{K}, \underbrace{0, \cdots, 0}_{L_\text{guard}}, \underbrace{x_{L_\text{data}}, \cdots, x_{2L_\text{data}-1}}_{L_\text{data}}, \underbrace{0, \cdots, 0}_{L_\text{guard}}}_{L},
  \cdots,
  \nonumber\\&
  \quad\,\,
  \underbrace{\psi^\mathcal{D}/\sqrt{K}, \underbrace{0, \cdots, 0}_{L_\text{guard}}, \underbrace{x_{L_\text{data} K}, \cdots, x_{L_\text{data}K-1}}_{L_\text{data}}, \underbrace{0, \cdots, 0}_{L_\text{guard}}}_{L}
  )^T
  \!
  .
  \label{eq:x_pilot_data}
\end{align}
When the power of the pilot symbol is identical to that of data symbols,
i.e.,~$E_\text{pilot} = K E_s$,
the resulting transmitted symbols with PSK signaling
containing embedded pilot symbol are equivalent to the original SC transmission
except that the guard symbols (i.e., null) of length~$L_\text{guard}$
are inserted before and after each pilot symbol~$\psi^\mathcal{D}/\sqrt{K}$.

\begin{figure}[tb]
  \centering
   \includegraphics[width = \hsize, clip]{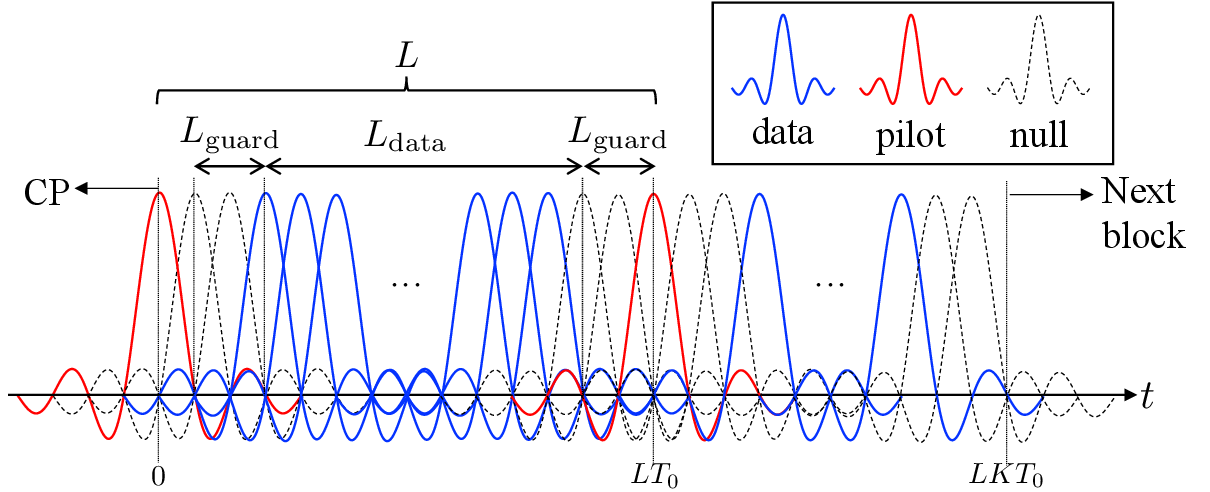}
   \caption{Time domain representation of SC signal with sinc pulses containing embedded pilot symbols for SC-DDE.}
  \label{signal}
\end{figure}

Fig.~\ref{signal} illustrates the transmitted SC signal~$s(t)$ with sinc pulses containing pilot symbols for SC-DDE,
where we consider the case with~$L_\text{guard} = 2$.
Here,
we set $E_\text{pilot} = K E_s$ such that
the power of the time domain pilot symbol in~\eqref{eq:pilot_time}
is the same as the average data symbol power,
i.e.,~$E_\text{pilot} = K E_s$.
It is apparent from Fig.~\ref{signal} that
the pilot symbol does not affect the peak power property of SC transmission.
On the other hand,
the insertion of the guard symbols causes average power reduction.
Hence,
the PAPR may increase as the number of guard symbols~$L_\text{guard}$ increases.

\section{Simulation Results}
\label{sec:simulation}

Through computer simulation,
we evaluate our proposed SC-DDE system
in terms of PAPR property and BER performance.
Unless otherwise noted,
we set the block length as~$N = 1024$.

\subsection{PAPR}

We first evaluate the PAPR property of SC-DDE compared to the conventional systems.
The PAPR is defined for each block of the continuous-time transmitted signal~$s \left( t \right)$ as
\begin{align}
  \text{PAPR}_{\ell}
  &=
  \frac{\displaystyle\max_{\ell T_\text{block} \leq t \leq \left( \ell + 1 \right) T_\text{block}} \left\{ \left| s \left( t \right) \right|^2 \right\}}{E \left\{ \left| s \left( t \right) \right|^2 \right\}},
\end{align}
where it is calculated without CP insertion in this subsection.
The PAPR is evaluated for~$s \left( t \right)$ by $8$ times oversampling with respect to the Nyquist interval~$T_0$.

\subsubsection{Without Channel Estimation}

\begin{figure}[tb]
  \centering
  \includegraphics[width = \hsize, clip]{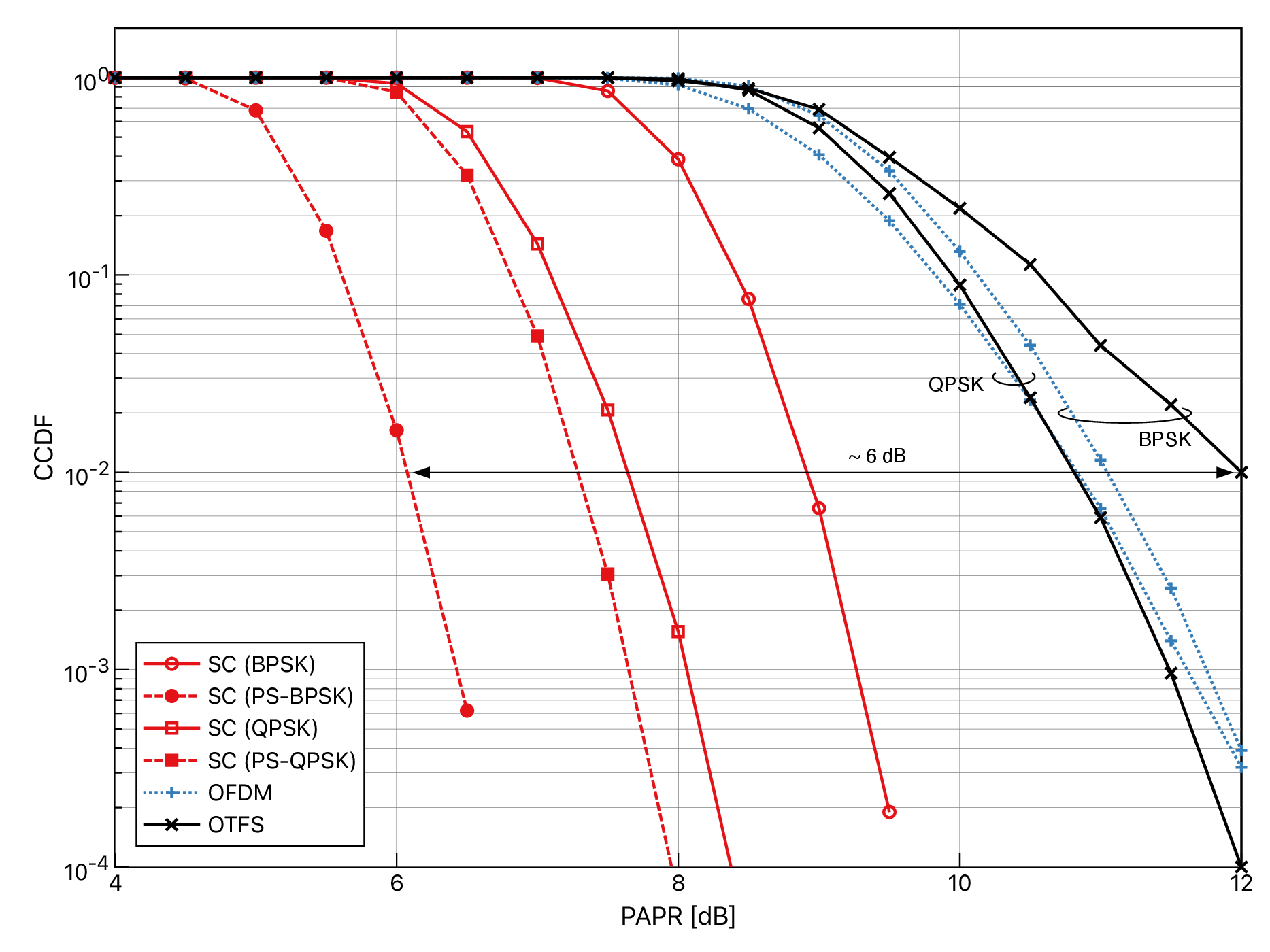}
  \caption{The PAPR of SC signal
    compared to OFDM and OTFS signals
    without embedded pilot symbol for channel estimation.
    ($N = 1024$, $L = K = 32$)
  }
  \label{papr}
\end{figure}

Fig.~\ref{papr} shows the PAPR characteristics of SC, OTFS, and OFDM signals in terms of complementary cumulative distribution function~(CCDF)
without pilot symbols for channel estimation.
In OTFS modulation,
we set $L = 32$ and $K = 32$.
Note that OFDM corresponds to the case with~$L = 1$ and $K = N = 1024$.
In Fig.~\ref{papr},
we evaluate BPSK and QPSK signaling for all schemes compared here.
We note that phase shifts of~$\pi/2$ and~$\pi/4$ are commonly applied to BPSK and QPSK in SC systems due to their lower PAPR property,
which are shown as phase shift BPSK~(PS-BPSK) and phase shift QPSK~(PS-QPSK) in Fig.~\ref{papr}, respectively.
In the case of OTFS and OFDM,
high PAPR is observed regardless of the modulation scheme,
whereas the PAPR of SC transmission depends on the modulation scheme.
In all the cases compared here,
SC transmission achieves significantly lower PAPR
compared to OTFS and OFDM,
and there is a PAPR gap of about $6$~dB evaluated at CCDF of~$10^{-3}$
between SC with PS-BPSK and OTFS with BPSK.


\subsubsection{With Channel Estimation}

\begin{figure}[tb]
  \centering
  \vspace{-3.2mm}
  \subfigure[BPSK]{\includegraphics[width=\hsize]{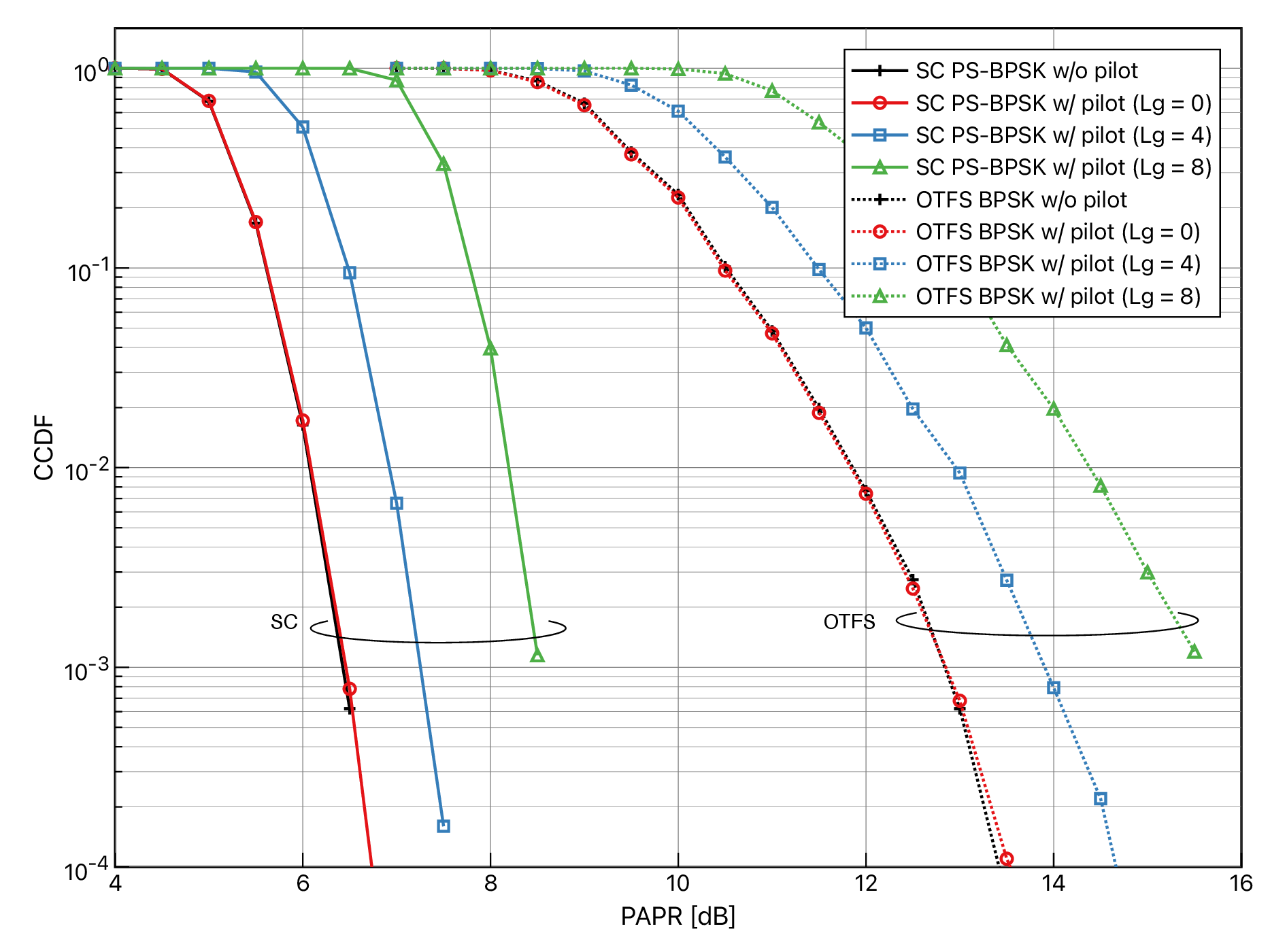}
    \vspace{-3.2mm}
    \label{papr_pilot_bpsk}}
  \subfigure[QPSK]{\includegraphics[width=\hsize]{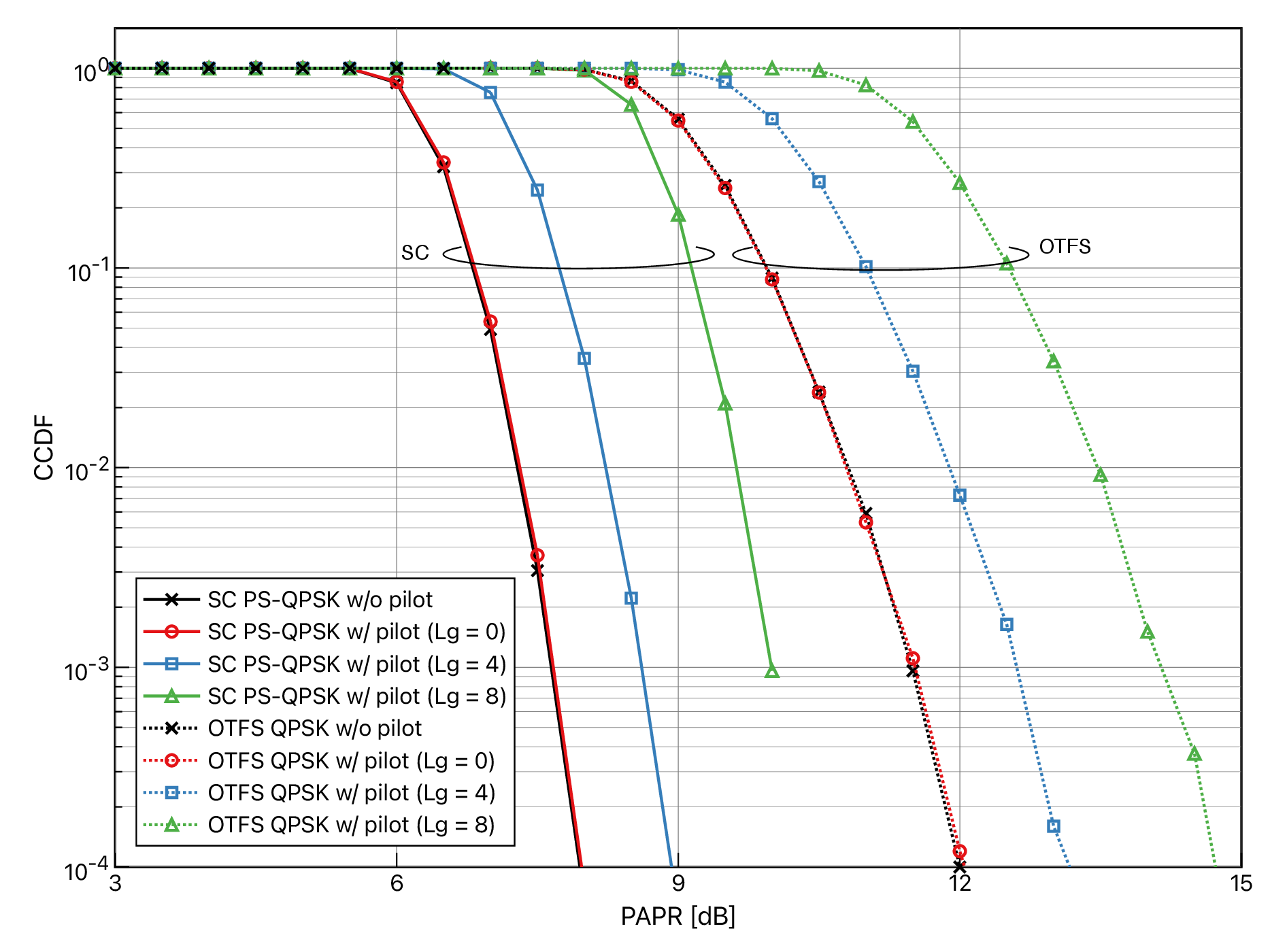}
    \vspace{-3.2mm}
    \label{papr_pilot_qpsk}}
  \caption{
    The PAPR comparison of SC and OTFS signals
    with embedded pilot symbol for channel estimation.
    ($N = 1024$, $L = K = 32$)
  }
  \label{papr_pilot}
\end{figure}



We next evaluate the PAPR characteristics of SC and OTFS
with embedded pilot symbols for channel estimation in Fig.~\ref{papr_pilot},
where we compare the BPSK case in Fig.~\ref{papr_pilot_bpsk} and the QPSK case in Fig.~\ref{papr_pilot_qpsk}.
We set the power of the embedded pilot symbol as~$E_\text{pilot} = K E_s$
such that the envelope of the time domain pilot symbol in~\eqref{eq:pilot_time}
is identical to that of PSK signaling.
In Fig.~\ref{papr_pilot}, we evaluate only the case with phase shift for SC transmission.
It should be noted that
the resource allocation of pilot and data symbols in the DD domain
is identical for SC-DDE and OTFS,
and the length of guard symbols is set as~$L_\text{guard} = 0, 4$, and~$8$.
It is apparent from Fig.~\ref{papr_pilot} that
the PAPR increases as the length of guard symbols~$L_\text{guard}$ increases.
Since the embedded pilot-aided channel estimation for SC-DDE introduced in Section~\ref{sec:channel_estimation}
does not affect the peak power property,
the resulting PAPR simply increases with the insertion of a total of $2L_\text{guard} K$ guard symbols, 
which leads to the average power reduction as illustrated in Fig.~\ref{signal}.
As a result,
SC is always superior to OTFS in terms of PAPR
compared under the same number of guard symbols~$L_\text{guard}$.
Furthermore, in Fig.~\ref{papr_pilot},
SC {\it with} embedded pilot symbol shows lower PAPR property than OTFS {\it without} pilot symbols.
Therefore,
SC-DDE is an effective approach in view of PAPR property.

\subsection{BER Performance}

We evaluate the error rate performance of SC-DDE through computer simulation
in terms of uncoded and coded BER.
For channel coding,
we employ a regular $(3,6)$ low-density parity-check~(LDPC) code with code rate~$R_c = 1/2$.
The channel is modeled as doubly-selective Rayleigh fading channel with
the number of paths~$P = 8$,
where each complex channel gain follows i.i.d. Rayleigh distribution with equal power profile,
i.e.,~$h_p \sim \mathcal{CN} \left( 0, 1/P \right)$.
In addition,
delay and Doppler taps corresponding to the $p$th path are fixed
as listed in Table~\ref{table:channel}
based on Fig.~2(a) of~\cite{surabhi2019low}.
In order to avoid the IBI,
the CP length is set as~$N_\text{CP} = 8$
so as to satisfy~$N_\text{CP} > l_\text{max} (= 7)$.

\begin{table}[tb]
  \begin{center}
    \caption{Delay and Doppler Parameters of Channel Model.}
    \begin{tabular}{c | c c c c c c c c}
      Path $p$ & 0 & 1 & 2 & 3 & 4 & 5 & 6 & 7
      \\ \hline \hline
      Delay tap~$l_{p} $ & 0 & 1 & 2 & 3 & 4 & 5 & 6 & 7 \\
      Doppler tap~$k_{p} $& 0 & 1 & 1 & 2 & 3 & 3 & 4 & 4
    \end{tabular}
    \label{table:channel}
  \end{center}
  \vspace{-3.0mm}
\end{table}

\subsubsection{Without Channel Estimation}

\begin{figure}[tb]
  \centering
  \includegraphics[width = \hsize, clip]{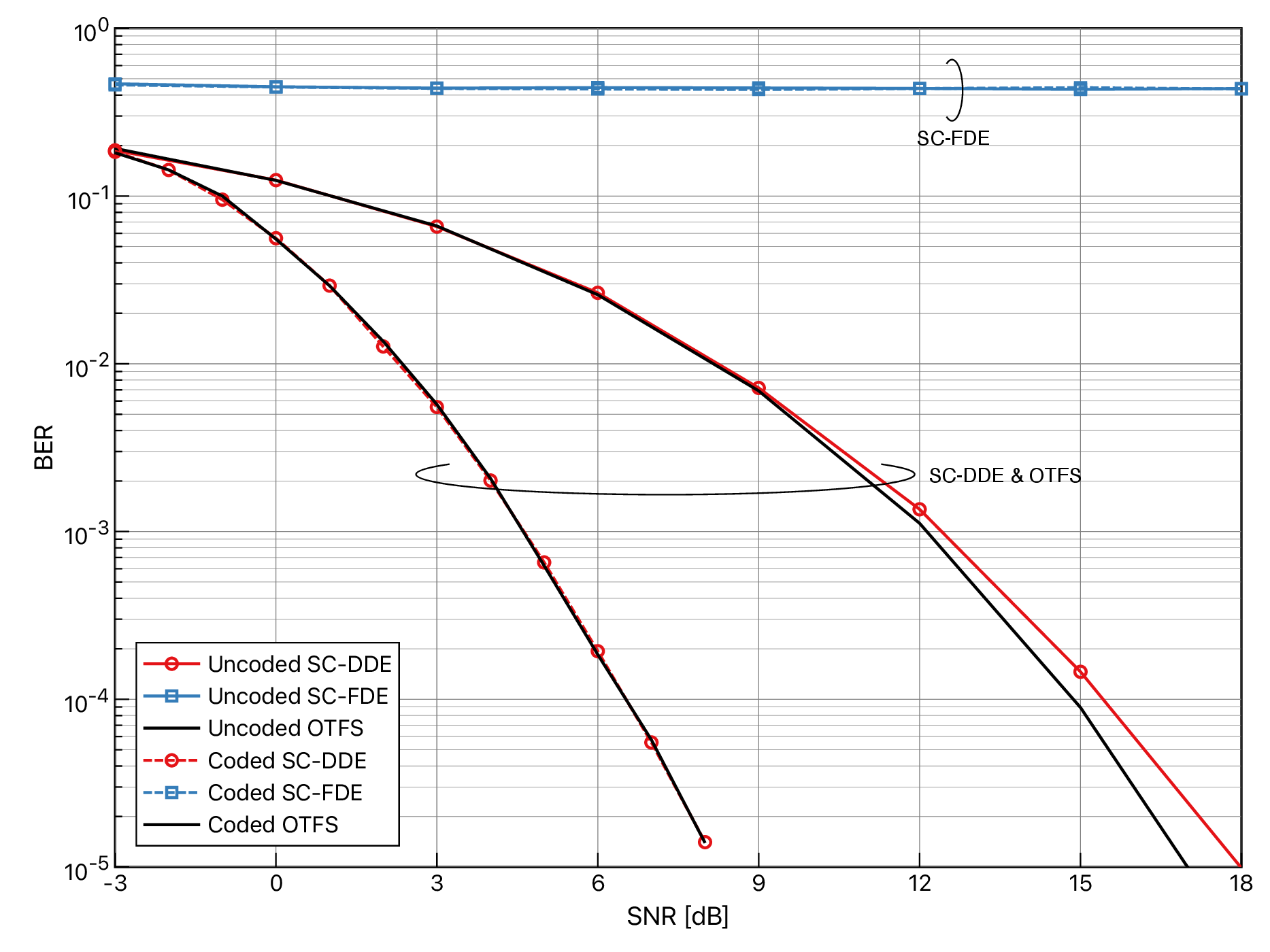}
  \caption{The uncoded and LDPC-coded BER performance of SC-DDE, SC-FDE, and OTFS with BPSK signaling over doubly-selective fading channel,
    where the receiver ideally knows the CSI without channel estimation.
    ($N = 1024$ and $L = K = 32$)}
  \label{ber}
\end{figure}

The uncoded and coded BER performance of SC-DDE is compared
with that of OTFS in Fig.~\ref{ber},
where the receiver employs DDE with the size of~$\left( L, K \right) = \left( 32, 32 \right)$
assuming the ideal CSI. 
Since BPSK modulation is performed for the block length of~$N = 1024$ in Fig.~\ref{ber},
the code length is given by~$1024$,
and the length of the information bit sequence is~$512$.
Each coded bit is mapped onto one BPSK symbol without embedded pilot symbol,
and the receiver employs DDE with the size of~$\left( L, K \right) = \left( 32, 32 \right)$.
For reference,
we also plot the results of SC-FDE~\cite{falconer2002frequency} in Fig.~\ref{ber}
as a performance metric of the conventional equalization approach for SC transmission.
In the conventional SC-FDE,
decoding fails to perform 
properly even with channel coding.
On the other hand,
the proposed SC-DDE achieves good BER performance
over doubly-selective channels.
Compared to OTFS,
the uncoded BER of SC-DDE is slightly inferior in high SNR region,
but this degradation should be negligible in terms of coded BER. 

\subsubsection{With Channel Estimation}

\begin{figure}[tb]
  \centering
  \includegraphics[width = \hsize, clip]{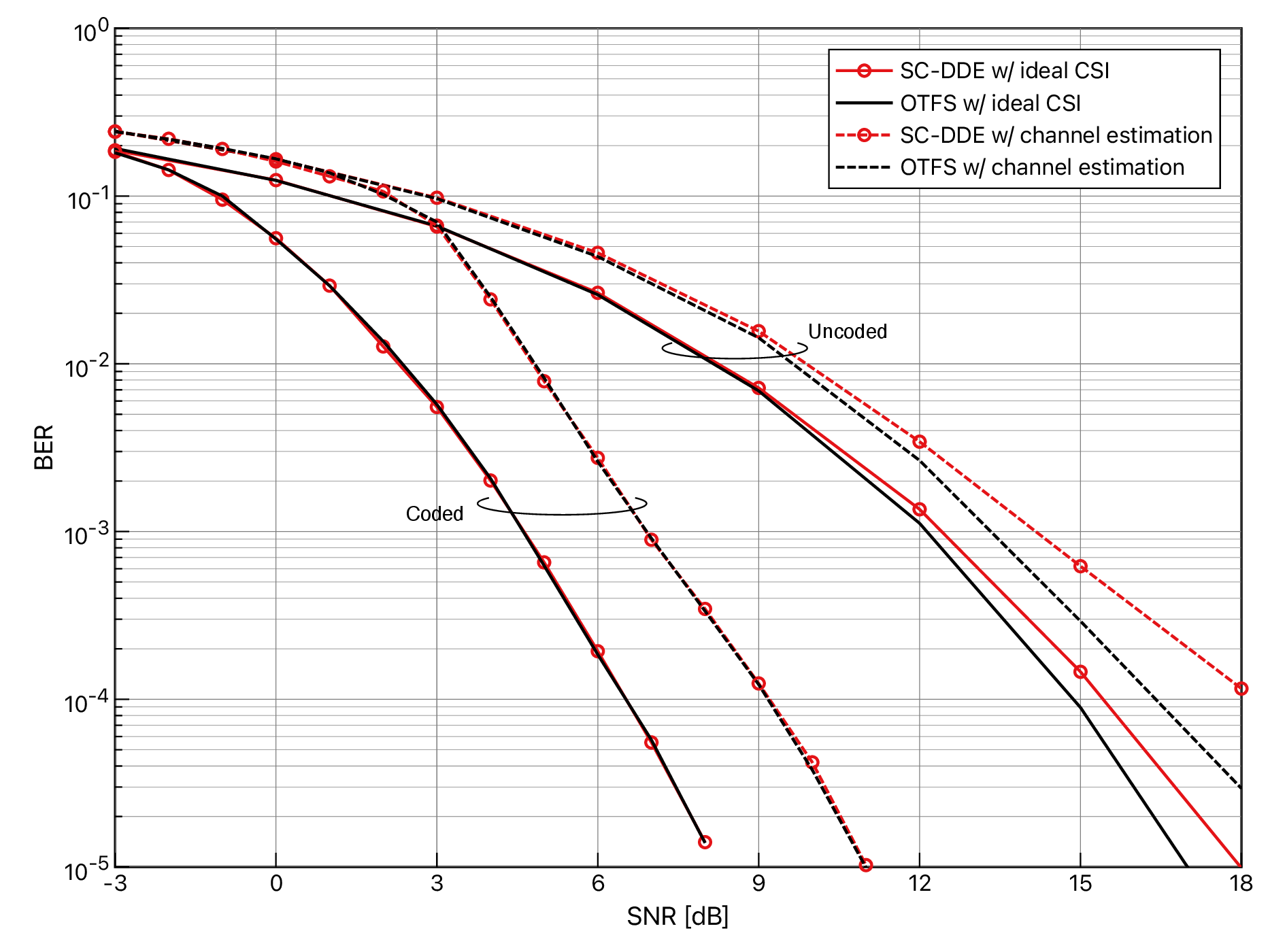}
  \caption{The uncoded and LDPC-coded BER performance of SC-DDE and OTFS with BPSK signaling over doubly-selective fading channel
    with embedded pilot-aided channel estimation.
    ($N = 1024$, $L = K = 32$, and~$L_\text{guard} = 7$)}
  \label{ber_pilot}
\end{figure}

We next compare the BER performance of SC-DDE and OTFS with embedded pilot-aided channel estimation
in Fig.~\ref{ber_pilot},
which are plotted by dashed lines.
Here,
we assume that the receiver knows delay and Doppler taps in Table~\ref{table:channel}
and estimates the corresponding complex channel gain~$h_p$ from the pilot symbol.
The number of guard symbols is set as~$L_\text{guard} = 7$ from~$L_\text{guard} = l_\text{max}$,
and thus,
the number of data BPSK symbols in each block is given by~$N_\text{data} = \left( L - 2L_\text{guard} - 1 \right) K = 544$.
Hence,
the code length with embedded channel estimation is set as~$544$,
and the length of the information bit sequence is given by~$272$.
The other parameters are identical to those in Fig.~\ref{ber},
and we also plot the results with ideal channel estimation by solid lines in Fig.~\ref{ber_pilot} for reference.
From the results,
the degradation of both SC-DDE and OTFS compared to the ideal channel estimation case is almost the same,
and SC-DDE is slightly inferior to OTFS in terms of uncoded BER.
Nevertheless, in this case as well,
the coded BER performance of our proposed SC-DDE is comparable with the conventional OTFS.

In conclusion,
the proposed SC-DDE is an effective approach for low PAPR transmission over doubly-selective channels
at the expense of high computational complexity at the receiver.

\section{Conclusion}
\label{sec:conclusion}

We have proposed delay-Doppler domain equalization~(DDE)
for single-carrier~(SC) transmission,
referred to as SC-DDE,
for doubly-selective fading channels.
To this end,
we first described the SC transmission over doubly-selective channels.
By introducing the discrete Zak transform~(DZT)
and its vectorization form as vectorized DZT~(VDZT),
we have introduced SC-DDE,
which enables simultaneous compensation of delay and Doppler shift.
Furthermore,
we have proposed the channel estimation scheme for DDE,
which does not affect the peak power property of SC transmission.
Through computer simulation,
our proposed SC-DDE significantly outperforms
the conventional SC-FDE in terms of the BER performance
at the expense of additional computational complexity for equalization and weight matrix calculation.
Compared with OTFS,
SC-DDE has much lower PAPR even with embedded pilot symbol for channel estimation,
while achieving almost the same BER performance
with channel coding.



\bibliographystyle{IEEEtran}
\bibliography{IEEEabrv,bibhama}

\begin{thebibliography}{10}

\bibitem{dahlman20134g}
E.~Dahlman, S.~Parkvall, and J.~Skold, {\em 4G: {LTE/LTE-advanced} for mobile
  broadband}.
\newblock Academic press, 2013.

\bibitem{ochiai2001distribution}
H.~Ochiai and H.~Imai, ``On the distribution of the peak-to-average power ratio
  in {OFDM} signals,'' {\em {IEEE} Trans. Commun.}, vol.~49, no.~2,
  pp.~282--289, 2001.

\bibitem{wang2006performance}
T.~Wang, J.~G. Proakis, E.~Masry, and J.~R. Zeidler, ``Performance degradation
  of {OFDM} systems due to {Doppler} spreading,'' {\em {IEEE} Trans. Wireless
  Commun.}, vol.~5, no.~6, pp.~1422--1432, 2006.

\bibitem{ochiai_instantaneous_2012}
H.~Ochiai, ``On instantaneous power distributions of single-carrier {FDMA}
  signals,'' {\em IEEE Wireless Commun. Lett.}, vol.~1, pp.~73--76, Apr. 2012.

\bibitem{myung2006single}
H.~G. Myung, J.~Lim, and D.~J. Goodman, ``Single carrier {FDMA} for uplink
  wireless transmission,'' {\em {IEEE} Veh. Technol. Mag.}, vol.~1, no.~3,
  pp.~30--38, 2006.

\bibitem{rikkinen2020thz}
K.~Rikkinen, P.~Kyosti, M.~E. Leinonen, M.~Berg, and A.~Parssinen, ``{THz}
  radio communication: {Link} budget analysis toward {6G},'' {\em {IEEE}
  Commun. Mag.}, vol.~58, no.~11, pp.~22--27, 2020.

\bibitem{falconer2002frequency}
D.~Falconer, S.~L. Ariyavisitakul, A.~Benyamin-Seeyar, and B.~Eidson,
  ``Frequency domain equalization for single-carrier broadband wireless
  systems,'' {\em {IEEE} Commun. Mag.}, vol.~40, no.~4, pp.~58--66, 2002.

\bibitem{hadani2017orthogonal}
R.~Hadani, S.~Rakib, M.~Tsatsanis, A.~Monk, A.~J. Goldsmith, A.~F. Molisch, and
  R.~Calderbank, ``Orthogonal time frequency space modulation,'' in {\em Proc.
  IEEE WCNC}, Mar. 2017.

\bibitem{lin2022orthogonal}
H.~Lin and J.~Yuan, ``Orthogonal {delay-Doppler} division multiplexing
  modulation,'' {\em IEEE Wireless Commun. Lett.}, vol.~21, no.~12,
  pp.~11024--11037, 2022.

\bibitem{wei2021orthogonal}
Z.~Wei, W.~Yuan, S.~Li, J.~Yuan, G.~Bharatula, R.~Hadani, and L.~Hanzo,
  ``Orthogonal time-frequency space modulation: A promising next-generation
  waveform,'' {\em IEEE Wireless Commun.}, vol.~28, no.~4, pp.~136--144, 2021.

\bibitem{li2021performance}
S.~Li, J.~Yuan, W.~Yuan, Z.~Wei, B.~Bai, and D.~W.~K. Ng, ``Performance
  analysis of coded {OTFS} systems over high-mobility channels,'' {\em {IEEE}
  Trans. Wireless Commun.}, vol.~20, no.~9, pp.~6033--6048, 2021.

\bibitem{lin2023multi}
H.~Lin, J.~Yuan, W.~Yu, J.~Wu, and L.~Hanzo, ``Multi-carrier modulation: An
  evolution from time-frequency domain to {delay-Doppler} domain,'' {\em arXiv
  preprint arXiv:2308.01802}, 2023.

\bibitem{raviteja2018practical}
P.~Raviteja, Y.~Hong, E.~Viterbo, and E.~Biglieri, ``Practical pulse-shaping
  waveforms for reduced-cyclic-prefix {OTFS},'' {\em {IEEE} Trans. Veh.
  Technol.}, vol.~68, no.~1, pp.~957--961, 2018.

\bibitem{yuan2021iterative}
Z.~Yuan, F.~Liu, W.~Yuan, Q.~Guo, Z.~Wang, and J.~Yuan, ``Iterative detection
  for orthogonal time frequency space modulation with unitary approximate
  message passing,'' {\em {IEEE} Trans. Wireless Commun.}, vol.~21, no.~2,
  pp.~714--725, 2021.

\bibitem{lampel2022otfs}
F.~Lampel, A.~Avarado, and F.~M. Willems, ``On {OTFS} using the discrete {Zak}
  transform,'' in {\em Proc. IEEE ICC Workshops}, May 2022.

\bibitem{surabhi2019peak}
G.~Surabhi, R.~M. Augustine, and A.~Chockalingam, ``Peak-to-average power ratio
  of {OTFS} modulation,'' {\em {IEEE} Commun. Lett.}, vol.~23, no.~6,
  pp.~999--1002, 2019.

\bibitem{wei2021charactering}
P.~Wei, Y.~Xiao, W.~Feng, N.~Ge, and M.~Xiao, ``Charactering the
  peak-to-average power ratio of {OTFS} signals: A large system analysis,''
  {\em {IEEE} Trans. Wireless Commun.}, vol.~21, no.~6, pp.~3705--3720, 2021.

\bibitem{gao2020peak}
S.~Gao and J.~Zheng, ``Peak-to-average power ratio reduction in pilot-embedded
  {OTFS} modulation through iterative clipping and filtering,'' {\em {IEEE}
  Commun. Lett.}, vol.~24, no.~9, pp.~2055--2059, 2020.

\bibitem{naveen2020peak}
C.~Naveen and V.~Sudha, ``Peak-to-average power ratio reduction in otfs
  modulation using companding technique,'' in {\em 2020 5th international
  conference on devices, circuits and systems (ICDCS)}, pp.~140--143, IEEE,
  2020.

\bibitem{hama2022noma}
Y.~Hama and H.~Ochiai, ``Time-frequency domain non-orthogonal multiple access
  for power efficient communications,'' {\em {IEEE} Trans. Wireless Commun.},
  vol.~22, pp.~5711--5724, Sept. 2023.

\bibitem{hama2023achievable}
Y.~Hama and H.~Ochiai, ``On the achievable spectral efficiency of
  non-orthogonal frequency division multiplexing,'' {\em {IEEE} Trans.
  Commun.}, vol.~71, pp.~6246--6257, Nov. 2023.

\bibitem{hlawatsch2011wireless}
F.~Hlawatsch and G.~Matz, {\em Wireless communications over rapidly
  time-varying channels}.
\newblock Academic press, 2011.

\bibitem{mohammed2021derivation}
S.~K. Mohammed, ``Derivation of {OTFS} modulation from first principles,'' {\em
  {IEEE} Trans. Veh. Technol.}, vol.~70, no.~8, pp.~7619--7636, 2021.

\bibitem{bolcskei1997discrete}
H.~Bolcskei and F.~Hlawatsch, ``Discrete {Zak} transforms, polyphase
  transforms, and applications,'' {\em {IEEE} Trans. Signal Process.}, vol.~45,
  no.~4, pp.~851--866, 1997.

\bibitem{nimr2018extended}
A.~Nimr, M.~Chafii, M.~Matthe, and G.~Fettweis, ``Extended {GFDM} framework:
  {OTFS} and {GFDM} comparison,'' in {\em Proc. IEEE GLOBECOM}, Dec. 2018.

\bibitem{surabhi2019low}
G.~Surabhi and A.~Chockalingam, ``Low-complexity linear equalization for {OTFS}
  modulation,'' {\em {IEEE} Commun. Lett.}, vol.~24, no.~2, pp.~330--334, 2019.

\bibitem{raviteja2019embedded}
P.~Raviteja, K.~T. Phan, and Y.~Hong, ``Embedded pilot-aided channel estimation
  for {OTFS} in {delay-Doppler} channels,'' {\em {IEEE} Trans. Veh. Technol.},
  vol.~68, no.~5, pp.~4906--4917, 2019.

\end{thebibliography}
\bstctlcite{BSTcontrol}

\end{document}